\documentclass[a4paper,oneside,english,british,traditabstract,usenatbib]{aa}
\usepackage[T1]{fontenc}
\usepackage[latin9]{inputenc}
\usepackage{babel}
\usepackage{array}
\usepackage{prettyref}
\usepackage{float}
\usepackage{multirow}
\usepackage{amsmath}
\usepackage{amssymb}
\usepackage{graphicx}
\usepackage{setspace}
\usepackage[authoryear]{natbib}
\usepackage[unicode=true,pdfusetitle,
 bookmarks=true,bookmarksnumbered=false,bookmarksopen=false,
 breaklinks=false,pdfborder={0 0 1},backref=false,colorlinks=false]
 {hyperref}
\usepackage{breakurl}

\makeatletter

\providecommand{\tabularnewline}{\\}

\usepackage{txfonts}
\@ifundefined{definecolor}
 {\usepackage{color}}{}

\bibpunct{(}{)}{;}{a}{}{,} 
\authorrunning{S. Sengupta et al.}
\titlerunning{A nova re-accretion model for J-type carbon stars}

\makeatother

\begin{document}

\title{A nova re-accretion model for J-type carbon stars}

\author{S. Sengupta$\dagger$\and R. G. Izzard \and H.H.B. Lau}

\institute{Argelander-Institut für Astronomie, Auf dem Hügel 71, D-53121 Bonn,
Germany.}

\mail{$\dagger$ sutirtha@astro.uni-bonn.de}

\date{Received August 1, 2013; Accepted October 3, 2013}

\abstract{The J-type carbon (J)-stars constitute $10-15\%$ of the observed
carbon stars in both our Galaxy and the Large Magellanic Cloud (LMC).
They are characterized by strong {\normalsize $\mathrm{^{13}C}$}
absorption bands with low {\normalsize $^{12}\mathrm{C}/^{13}\mathrm{C}$}
ratios along with other chemical signatures peculiar for typical carbon
stars, e.g. a lack of s-process enhancement. Most of the J-stars are
dimmer than the N-type carbon stars some of which, by hot-bottom burning,
make {\normalsize $\mathrm{^{13}C}$} only in a narrow range of masses.
We investigate a binary-star formation channel for J-stars involving
re-accretion of carbon-rich nova ejecta on main-sequence companions
to low-mass carbon-oxygen white-dwarfs. The subsequent evolution of
the companion stars in such systems is studied with a rapid binary
evolutionary code to predict chemical signatures of nova pollution
in systems which merge into giant single stars. A detailed population
synthesis study is performed to estimate the number of these mergers
and compare their properties with observed J-stars. Our results predict
that such nova polluted mergers evolve with low luminosities as well
as low{\normalsize{} $^{12}\mathrm{C}/^{13}\mathrm{C}$} ratios like
the majority of observed J-stars (e.g. in the LMC) but cannot account
for the observed fraction of J-stars in existing surveys of carbon
stars. }

\keywords{Stars: carbon -- Stars: chemically peculiar -- (Stars:) novae, cataclysmic
variables -- (Stars:) white dwarfs -- Stars: abundances -- Stars:
statistics}

\maketitle

\section{Introduction}

Stars spend about 90\% of their lifetime on the main sequence, during
which they convert hydrogen (H) to helium (He) in their cores by the
pp-chain or the CNO-cycle. Following core hydrogen exhaustion they
leave the main sequence and depending on their mass, many proceed
to advanced stages of nuclear burning. In low and intermediate-mass
stars ($M\sim0.8-8M{}_{\odot}$), the triple-$\alpha$ reaction acts
as a source of energy after depletion of internal hydrogen and produces
carbon during their He-burning phase of evolution. When these stars
ascend the asymptotic giant branch (AGB), the products of He-burning
(mainly $\mathrm{^{12}C}$) along with s-process elements formed through
neutron captures are mixed into the convective envelope during the
thermally pulsing (TP) phase and brought to the stellar surface by
third dredge-up (TDU). This AGB scenario explains the origin of the
majority of all carbon (C)-stars (characterized by surface $\mathrm{C/O>1}$
by number) - the N-type stars \citep{1983ARA&A..21..271I}. However,
spectral classification of C-stars reveals other peculiar scenarios
of carbon enrichment especially in binaries (\citealp{1998ARA&A..36..369W}).
Mass transfer from a carbon-rich AGB primary star can pollute its
binary companion with enough carbon to turn it into a dwarf carbon
star which explains the origin of the CH and Ba-stars \citep{1998A&A...332..877J}.
In this work, we investigate a binary model that aims to explain the
origin of possibly the least understood class among all C-stars -
the J-type (J)-stars - whose evolutionary origin has remained a mystery
for decades.

\subsection{J-stars}

The J-stars are a class of red-giant C-stars that are characterized
by strong $\mathrm{^{13}C}$ bands (\citealt{1954AnAp...17..104B})
and low $\mathrm{^{12}C/^{13}C}$ ratios (\citealp{1999A&A...345..233O}),
and constitute a significant fraction ($10-15\%$) of all carbon stars
in our Galaxy (\citealt{2000ApJ...536..438A}) and the LMC (\citealt{2003MNRAS.341..534M}).
AGB models of C-stars cannot explain most of the chemical peculiarities
associated with J-stars e.g. no s-process overabundance and Li enhancements
(\citealt{2003MNRAS.341.1290H}). Moreover, their luminosities and
variability classes indicate that they are less evolved objects than
the N-type C-stars \citep{2000ApJ...536..438A}. However, the presence
of high luminosity J-stars in our Galaxy (e.g. WZ Cas) suggests the
existence of at least two types of J-stars, with different formation
scenarios. Moreover, an intriguing sub-class of J-stars known as silicate
carbon stars, characterized by oxygen-rich circumstellar material
possibly stored in a circumbinary disk (\citealt{1991MNRAS.249..409L}),
points to a binary origin for this peculiar class of C-stars (e.g.
BM-Gem; \citealt{2008A&A...490..173O}, \citealt{2008ApJ...682..499I}).
A recent study exploring white-dwarf (WD) red-giant (RG) binary mergers
in the context of the R-type C-stars, has found a possible evolutionary
channel for J-stars \citep{2013MNRAS.430.2113Z}. We investigate another
binary scenario considering chemical pollution of main-sequence (MS)
companions to WDs in classical novae explosions and study their long-term
evolutionary outcome in context of peculiarities associated with J-stars.

\subsection{$\mathrm{^{13}C}$ enhancement scenarios}

The triple-$\alpha$ process operating in cores of AGB stars makes
$\mathrm{^{12}C}$ which is subsequently brought to their surface
during dredge up thus forming a C-star. The synthesis of $\mathrm{^{13}C}$
requires a further proton capture on $\mathrm{^{12}C}$ that can only
occur in regions with conditions suitable for operation of the CNO
cycle. The $\beta$-unstable isotope $\mathrm{^{13}N}$ produced in
CN cycle by $\mathrm{^{12}C(p,\gamma)^{13}N}$ decays to $\mathrm{^{13}C}$
which when transported to the stellar surface leads to low $\mathrm{^{12}C/^{13}C}$
ratios characteristic of J-stars. Such processes occur during the
evolutionary stages of both single and binary stellar systems as described
next.

\subsubsection{Single stars}

It is not easy to construct a model of an AGB star with the peculiar
properties of J-stars. AGB star models predict low $\mathrm{^{12}C/^{13}C}$
ratios for masses $M>4M{}_{\odot}$ if temperatures at the base of
the convective envelope are sufficiently high for hot bottom burning
(HBB, \citealt{1995ApJ...442L..21B}). However, the CN-cycle operating
under such conditions destroys $\mathrm{^{12}C}$ and consequently
lowers the $\mathrm{C/O}$ ratio in the envelope to below $1$ making
the star $\mathrm{\mathrm{O}}$-rich. Only narrow ranges of masses
and mass-loss rates lead to short phases ($\sim0.1\,{\rm Myr}$) on
the AGB with both $\mathrm{C/O>1}$ and $\mathrm{^{12}C/^{13}C}<10$
corresponding to J-stars. Moreover, TDU in the TPAGB phase also enriches
of the envelope with s-process elements that are synthesized in the
inter-pulse period when $\mathrm{^{13}C}$ formed in the intershell
region acts as a neutron source via the $\mathrm{^{13}C}(\alpha,n)\mathrm{^{16}O}$
reaction (\citealt{1995ApJ...440L..85S}). However, observational
studies of J-stars show that abundances of s-process elements with
respect to iron for most of them are nearly solar (e.g. \citealt{1985ASSL..114..243U}).
Later works also conclude that mean heavy element abundance among
the J-stars in their sample is compatible with no enrichment \citep{2000ApJ...536..438A}. 

Statistically the narrow range of masses for which the HBB AGB models
predict such chemical features cannot account for the observed fraction
of J-stars among C-stars . Luminosity estimates from carbon-star surveys
in the LMC also suggest that the HBB AGB models cannot explain the
dimmer ($M\mathrm{_{Bol}>-5}$) majority of J-stars (\citealt{2003MNRAS.341.1290H}).

\subsubsection{Binary systems}

While the HBB AGB scenario explains qualitatively the luminous J stars,
a different chain of events seems to be responsible for the anomalous
features observed in the majority of J-stars. Some other classes of
C-stars also show traits common with J-stars e.g. the R and CH stars
which also show low $\mathrm{^{12}C/^{13}C}$ ratios (\citealt{1992AJ....103.2035V})
and are associated with binary formation scenarios (e.g.\citealt{2007A&A...470..661I}).
Moreover observations of silicate carbon stars which constitute about
$10\%$ of J-stars (\citealt{1991MNRAS.249..409L}), provide further
evidence of a link with binarity (\citealt{1990AJ.....99.1612L}).
Chemical pollution in mass-transfer phases in binary systems during
a nova-phase (\citealt{1999MNRAS.309..245S}) can result in peculiar
surface abundances for the secondary star in a semi-detached state
(\citealt{1997MNRAS.290..283M}). Such pollution in nova systems presents
a case of particular interest in the context of the observed $\mathrm{^{13}C}$
enhancements in J-stars.

\subsection{Classical novae}

Novae are thermonuclear explosions in close binaries of the cataclysmic
variable (CV) type (\citealt{1995CAS....28.....W}) with a WD accreting
hydrogen-rich matter from a companion star undergoing Roche lobe overflow
(RLOF). Among them, classical novae involve RLOF from a MS star to
a WD (CO or ONe) in a close binary system resulting from a preceding
common envelope (CE) phase (\citealt{1976IAUS...73...75P}). At low
accretion rates (below $10^{-7}\, M{}_{\odot}\mathrm{yr^{-1}})$,
hydrogen is compressed to degenerate conditions until ignition occurs,
which leads to a thermonuclear runaway (TNR). Explosive hydrogen burning
synthesizes some short-lived $\beta$-unstable nuclei (e.g.$\mathrm{^{13}N}$,
$\mathrm{^{15}O}$) which are transported by convection to the cooler
outer envelope where isotopes like $\mathrm{^{13}C}$ are produced
from subsequent $\beta$-decays. These decays release sufficient energy
which causes an explosive outburst in the outer shells of the WD accompanied
by mass ejection with typical velocities $10^{2}-10^{3}$ km s$^{-1}$
(\citealt{1998PASP..110....3G}). 

The energetics of the eruption, including the speed and mass of the
ejecta, as well as the ejecta composition, are determined by the mass
of the underlying WD, $M\mathrm{_{WD}}$, and the mass accretion rate,
$\dot{M}$. In classical novae, $\dot{M}$ typically ranges between
$10^{-10}-10^{-8}\, M{}_{\odot}\mathrm{yr^{-1}}$ (\citealp{2007ApJ...663.1269N}).
For\textbf{ $\dot{M}=\frac{d\dot{M_{stable}}}{dt}\sim10^{-7}\, M{}_{\odot}\mathrm{yr^{-1}}$},
the accreted material can burn steadily resulting in growth of the
white dwarf mass possibly up to the Chandrasekhar limit leading to
a Type Ia supernova (\citealt{1973ApJ...186.1007W}, \citealt{1992IAUS..151..225W},
\citealt{1996ApJ...470L..97H}).

\subsection{Nova nucleosynthesis}

Observations of nova ejecta often show C, N and O overabundances with
respect to solar indicating that there is some mixing between the
core and the accreted envelope (\citealt{1998PASP..110....3G}). The
mixing between the accreted envelope and the underlying white dwarf
is a necessary condition both to power the explosion and to interpret
the extra-solar metallicities observed in nova ejecta (\citealt{2001coev.conf..149H}).
Even though the exact mechanism and the extent of this mixing remain
unclear, the WD core composition plays a crucial role in the subsequent
nucleosynthesis making it essential to distinguish between novae occurring
on CO and ONe WDs (\citealt{2012MNRAS.427.2411G}) .

\subsubsection{$^{13}$C production in novae}

The synthesis of $^{13}$C in novae is initiated through $\mathrm{^{12}C(p,\gamma)^{13}N}$
when the temperature at the base of the accreted material on the WD
surface reaches about $10^{7}\mbox{K}$ as required for the cold CNO
cycle to operate. The fate of $^{13}$C is determined by the competition
between destruction via $\mathrm{^{13}C(p,\gamma)^{14}N}$ which operates
near the burning shell, and production by means of $\mathrm{^{13}N(\beta^{+})^{13}C}$
in the outer, cooler layers of the envelope where a fraction of $\mathrm{^{13}N}$
is carried to via convection just above the burning shell. The \citet{1998ApJ...494..680J}
nova models show much higher overproduction of $^{13}$C for CO novae
as compared to ONe novae. This is expected primarily because of the
higher initial $\mathrm{^{12}C}$ content of CO WDs, and the operation
of the hot CNO cycle when $\mathrm{^{13}N(p,\gamma)^{14}O}$ is faster
than $\mathrm{^{13}N(\beta^{+})^{13}C}$ (at base temperatures $\mathrm{>10^{8}}\mbox{K}$
) in more massive ONe WDs.

All of these model calculations assume a uniform WD composition without
explicitly taking into account predictions from detailed evolutionary
calculations. The nova ejecta composition depends critically on the
initial composition of the WD material mixed with the accreted matter
during explosion, and the predicted yields could well show larger
enhancements of heavier (CNO) elements if outer layers of the underlying
WD has a non-uniform composition(e.g. $\mathrm{C:O=50:50}$ for a
CO nova). This could also account for the wide range of $\mathrm{C/O}$
observed in nova ejecta (\citealt{1998PASP..110....3G}). \citet{2004ApJ...612..414J}
refer to a $0.6\, M_{\odot}$ CO WD model with initial $\mathrm{C}:\mathrm{O}=60:40$
that gives ejecta with $\mathrm{C/O>1}$. Such a case is of particular
interest in context of carbon ($\mathrm{^{13}C}$ in particular) pollution
in nova explosions due to re-accretion of ejecta from the WD on its
companion, and is considered in a systematic approach as detailed
in the following section.

\pagebreak{}

\section{Method}

\label{sec:Method}This section outlines the nova re-accretion model
we use to investigate pollution in MS companions to WDs (WD-companions
hereafter) in nova binaries. The input physics and assumptions used
in \textit{binary\_c/nucsyn} -- our population nucleosynthesis code
based on the \textit{BSE} code of \citet{2002MNRAS_329_897H}, extended
to include various nucleosynthesis algorithms \citep{Izzard_et_al_2003b_AGBs,2006A&A...460..565I,2009A&A...508.1359I}
-- are also described.

\subsection{Nova re-accretion model}

\label{sub:The-Nova-Re-accretion}We use a simple geometric prescription
to estimate the fraction of ejected mass re-accreted by the WD-companion
(secondary) during the nova-outbursts following the approach used
to study chemical pollution of the secondary during novae by \citet{1994MNRAS.268..749S}.
Though nova shells are often observed to be asymmetric, this is most
likely caused by the interaction of the expanding shells with the
secondary star \citep{1995CAS....28.....W}. Hence, in order to estimate
the amount of ejected material that is accreted back on the surface
of the secondary, it is justifiable to assume a spherically symmetric
outburst with a geometric factor for the fraction of the nova-ejecta
intercepted by the secondary,

\begin{equation}
f_{\mathrm{nova}}=\frac{\pi R_{2}^{2}}{4\pi a^{2}}\,,\label{back accretion facor}
\end{equation}
which is the ratio of the cross-sectional area of the secondary star
of radius $R{}_{2}$ to the area of a sphere of radius equal to the
binary separation $a$. The mass of material re-accreted by the secondary
star is thus given by, 
\begin{equation}
\Delta M_{\mathrm{re-acc}}=\left(\frac{R_{2}}{2a}\right)^{2}\Delta M_{\mathrm{ej}}\,,\label{mass re-accreted}
\end{equation}
where $\Delta M_{\mathrm{ej}}$ is the mass ejected during the outburst.
During the nova (semi-detached) phase when the secondary overfills
its Roche lobe, $R\mathrm{_{2}}$ is approximately the Roche lobe
radius $R\mathrm{_{L,2}}$, and hence Eq.~\ref{back accretion facor}
gives the fraction $f_{{\rm nova}}$ as a analytic function of the
mass ratio.

Eq.~\ref{mass re-accreted} provides an upper estimate for the amount
of material re-accreted, assuming all of the nova ejecta intercepted
by the secondary sticks on to its surface. Detailed (3D) hydrodynamical
studies also predict significant re-accretion of the ejecta in nova
explosions on the surface of the WD companion (\citealt{2010nuco.confE.203C})
and justify our simple choice for the prescription for re-accretion
also used in previous works investigating similar effects of nova
pollution (e.g. \citealt{1997MNRAS.290..283M}).

\subsection{Nova input physics}

During novae, we assume that all but a small fraction $\mathrm{\epsilon}=0.001$
of the accreted material is ejected in the explosion so that the WD
does not grow significantly (\citealt{2007MNRAS.374.1449E}). \citet{1998ApJ...494..680J}
yields are used for the composition of the ejecta of classical novae
with WD masses in the range of $0.8-1.35\, M_{\odot}$ including both
CO \& ONe WD models, along with an additional $0.6\, M_{\odot}$ CO-WD
explosion model provided by J. Jose (private communication). 

\textbf{}
\begin{figure}
\centering{}\textbf{\includegraphics[angle=270,scale=0.35]{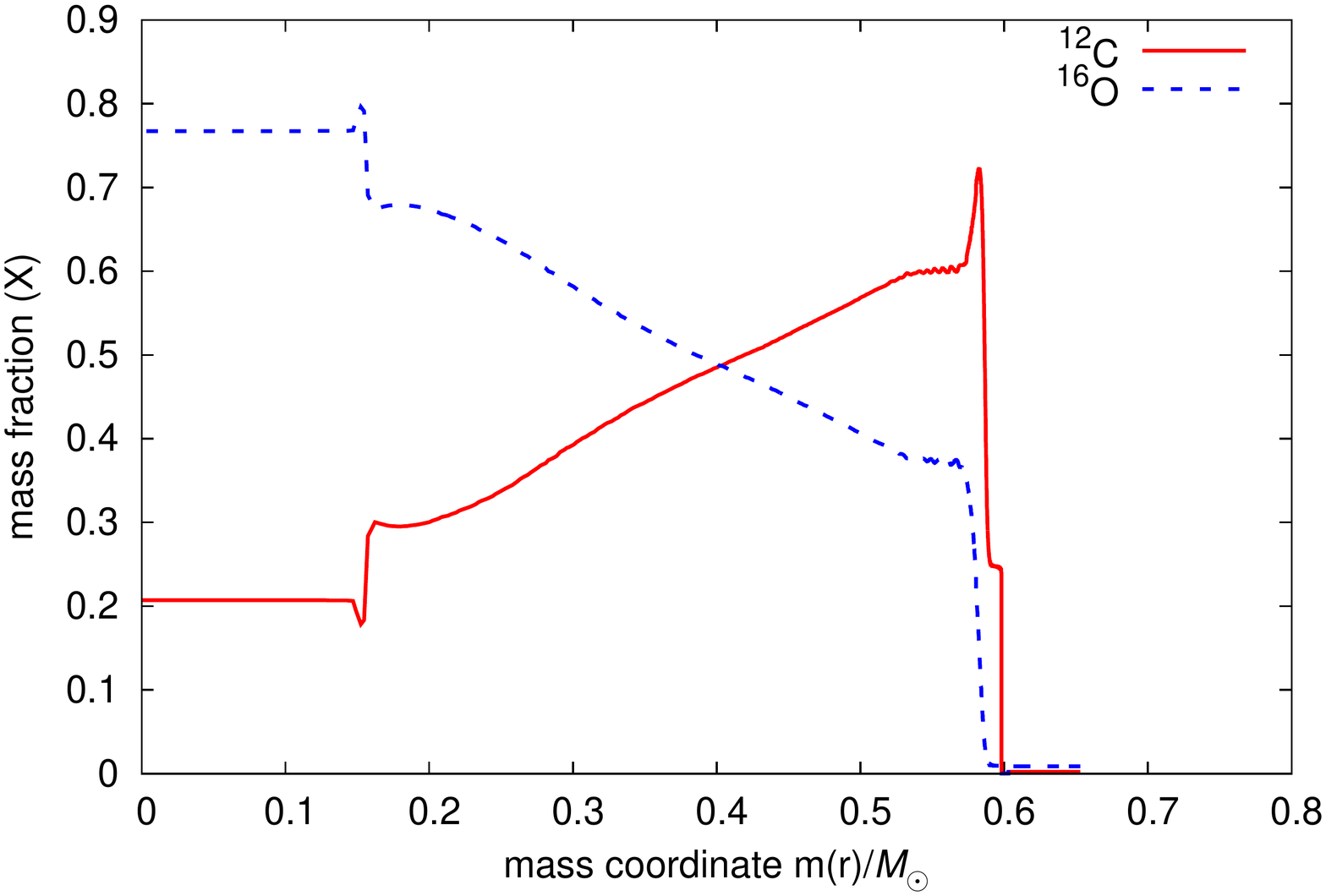}}\caption{\textbf{\label{fig:Internal-composition-of}}Mass fractions of carbon
and oxygen in the core of a TPAGB star with initial $M=3\, M{}_{\odot}$
and $Z=0.02$ obtained using the MESA code (\citealt{2013ApJS..208....4P})
illustrating that the outermost layers with mass coordinate $m(r)\gtrsim0.5\, M_{\odot}$
of a $0.6\, M_{\odot}$ CO-WD are expected to have $\mathrm{C}/\mathrm{O}>1$
(by number).}
\end{figure}

All the CO-WD nova models assume a uniform composition ${\rm C:O=50:50}$
as initial input for the nucleosynthesis calculations in the explosion.
However, a low mass CO-WD is expected to be carbon-rich ($\mathrm{C/O}>1$)
in its outer layers that are mixed into the accreted envelope during
the nova explosion (cf. Fig.~\ref{fig:Internal-composition-of}).
To account for this, an enhanced ${\rm C/O}\sim2$ is adopted for
the ejecta composition of this $0.6\, M_{\odot}$ CO-WD model. This
choice is well within the observed range of ejecta compositions for
classical novae as listed in the compilation of \citet{1998PASP..110....3G},
with even higher ${\rm C/O}$ for the ejecta of CO-novae e.g. V842
Cen with a estimated ${\rm C/O}\sim5$ (\citealt{1994A&A...291..869A}).

\subsubsection{Common envelope prescription}

The evolution of all CV systems, including those that go through a
nova phase, necessarily involves a common envelope (CE) phase which
shrinks the orbital separation allowing a subsequent semi-detached
phase of mass-transfer from the secondary (e.g. a MS star for classical
novae). Hence, the CE phase plays an important role in determining
which binary systems go through a nova phase after the envelope is
ejected. We treat the CE phase using the $\alpha$-formalism of \citet{1976IAUS...73...75P}
assuming that the orbital energy release during the inspiral is transferred
to the envelope with an efficiency $\mathrm{\alpha_{CE}}$, and can
expand and eject at least some of the envelope \citep{2002MNRAS_329_897H}.
The envelope binding energy is parametrized as,

\begin{equation}
E{}_{\mathrm{bind}}=-\frac{GM_{\mathrm{env}}M_{\mathrm{c}}}{\lambda R}\,,
\end{equation}
where $M_{\mathrm{env}},\, M_{\mathrm{c}}$ and $R$ are the envelope
mass, core mass and radius of the star and $\lambda$ is a measure
of central condensation of the star that is calculated from detailed
stellar models (\citealt{2000A&A...360.1043D}). The choice of the
parameter $\mathrm{\alpha_{CE}}$ is critical for any population synthesis
study involving nova binaries \citep{2004ApJ...602..938N}. It is
most probably not a constant (\citealt{2012MNRAS.419..287D}) though
population synthesis models usually assume a constant $\mathrm{\alpha{}_{CE}<1}$.
Our systematic approach to constrain $\alpha\mathrm{_{CE}}$ for nova
systems, as outlined in Appendix~A, involves estimating the expected
rate of novae in our Galaxy ($Z=0.02$) in order to compare with existing
observational constraints on the Galactic nova rate (\citealt{2002AIPC..637..462S}).
Based on the results of our calculations (cf. Table~\ref{tab:Dependence-of-novae})
we adopt a constant $\alpha\mathrm{_{CE}}=0.2$ for our population
synthesis work which is also consistent with recent observational
studies of post-CE binaries constraining $\alpha\mathrm{_{CE}}$ (\citealt{2010A&A...520A..86Z},
\citealt{2011MNRAS.411.2277D}), and briefly discuss the systematic
effect of this choice on our population synthesis predictions.

\subsection{binary\_c/nucsyn}

To investigate the re-accretion of nova ejecta by WD-companions according
to the prescription described in Section~\ref{sub:The-Nova-Re-accretion},
the \emph{binary\_c/nucsyn} code is used to evolve grids of binaries
at metallicities $Z=0.008\mbox{ and }0.02$ for a range of initial
primary ($M\mathrm{_{1}}$) and secondary ($M\mathrm{_{2}}$) masses
with $M\mathrm{_{1}}>M\mathrm{_{2}}$, and initial separations ($a$)
which give nova systems within a Hubble time ($\mathrm{13.7}\mathrm{\mbox{Gyr}}$).
The following limits are therefore chosen for $M\mathrm{_{1}}$, $M\mathrm{_{2}}$
and $a$,

\begin{align}
 & M\mathrm{_{1}}:0.8-6\, M_{\odot},\nonumber \\
 & M\mathrm{_{2}}:0.1\, M_{\odot}-M_{1}\mbox{ and}\\
 & a:10-10^{4}\, R_{\odot}\,.\nonumber 
\end{align}

\subsubsection{Initial distributions}

The initial distributions of binary parameters $M\mathrm{_{1}}$,
$M\mathrm{_{2}}$ and $a$ are given by the functions $\Psi(\ln M_{1})$,
$\Phi(\ln M_{2})$ and $\chi(\ln a)$ with,

\begin{equation}
\Psi(\ln M)=M\xi(M)\,,\label{primary mass dist.}
\end{equation}
where $\xi(M)$ is the initial mass function (IMF) of \citet{KTG1993MNRAS-262-545K}.
The distribution $\Phi$ for $M\mathrm{_{2}}$ is chosen to be flat
in the mass ratio $q=M_{2}/M_{1}$ and the separation distribution
$\chi$ is taken to be flat in $\log\, a$, to facilitate comparison
with previous works of \citet{2010ApJ...720.1752P,2004ApJ...602..938N}.

\subsubsection{Binary fraction}

We assume a constant binary fraction $f{}_{\mathrm{bin}}=0.5$ which
corresponds to an equal number of single and binary systems in our
stellar population \citep{1991A&A...248..485D} .

\subsubsection{Binary grid}

With the above choices for initial binary parameters, a logarithmic
grid is set up in $M_{1}-M_{2}-a$ space for all binary stars and
in $M$ (initial mass with same choice of distribution and range as
$M_{1}$) for single stars. The grid is split into $n$ stars per
dimension such that each star represents the centre of a logarithmic
grid-cell of size $\delta V$ where,

\begin{equation}
\delta V=\left\{ \begin{array}{cc}
\delta\ln M & \textrm{single stars}\,,\\
\delta\ln M_{1}\,\delta\ln M_{2}\,\delta\ln a & \textrm{binary stars}\,,
\end{array}\right.\label{log space vol.}
\end{equation}
with
\begin{equation}
\delta\ln x=\frac{\ln x_{\textrm{max}}-\ln x_{\textrm{min}}}{n-1}\mathrm{\,,}\label{log spacing}
\end{equation}
where $x$ represents $M$, $M_{1}$, $M_{2}$ or $a$ and $x\mathrm{_{\textrm{min}}}$
and $x\mathrm{_{\textrm{max}}}$ are the grid limits. The probability
for the $i$$^{th}$ model in the grid is given by,

\begin{equation}
p_{i}=\left\{ \begin{array}{cc}
\Psi(\ln M_{i})\delta V & \textrm{single stars\,,}\\
\Psi(\ln M_{1i})\,\Phi(\ln M_{2i})\,\chi(\ln a_{i})\:\delta V & \textrm{binary stars}\,,
\end{array}\right.\label{eq:probability}
\end{equation}
where
\begin{equation}
\sum_{i}p_{i}=1\,,
\end{equation}
over the entire range of parameter space i.e. over masses (e.g. over
$M$ for single stars) and separations. Hence for binaries,

\begin{equation}
\sum_{i}={\displaystyle \sum_{M_{1}=0.8\,\mathrm{M}_{\odot}}^{M_{1}=6\,\mathrm{M}_{\odot}}}\sum_{M_{2}=0.1\,\mathrm{M}_{\odot}}^{M_{1}}\sum_{a=10\,\mathrm{R}_{\odot}}^{10^{4}\,\mathrm{R}_{\odot}}\mathrm{\,.}\label{eq:grid loops}
\end{equation}

\subsubsection{Population synthesis\label{sub:Population-Synthesis}}

The re-accretion model as outlined in Section~\ref{sub:The-Nova-Re-accretion}
is applied to grids of binary models at a fixed metallicities $Z=0.02$
(solar) and $0.008$ (LMC). The initial abundances are chosen to be
solar-scaled based on the values of \citet{1989GeCoA..53..197A}.
The number of stars of a certain type (e.g. a C or J-star) is counted
by calculating the probability of the existence of the stellar system
(single or binary) given by Eq.~\ref{eq:probability} and the time
it spends in the evolutionary phase of interest $\Delta t_{i}$ from,
\begin{align}
N & =\sum_{i}S\times p_{i}\Delta t_{i}\delta\times\left\{ \begin{array}{cc}
(1-f_{\mathrm{bin}}) & \textrm{single stars}\,,\\
f_{\mathrm{bin}} & \textrm{binary stars}\,,
\end{array}\right.\label{eq:counting}
\end{align}
where $\delta=1$ if the star is in the phase of interest (and zero
otherwise), the binary fraction $f_{\mathrm{bin}}=0.5=1-f_{\mathrm{bin}}$
and $S=7.086\mathrm{\,\mathrm{yr^{-1}}}$ is the (constant) star formation
rate obtained following the prescription of \citet{2002MNRAS_329_897H}.
Thus, for C-stars, $\mathrm{\delta=1}$ if $\mathrm{C/O}>1$, and
if in addition $\mathrm{\mathrm{^{12}C/^{13}C}<10}$, they also classify
as J-stars. Also, C and J-stars are only counted as giants (stellar
types GB, EAGB, TPAGB; \citealt{2002MNRAS_329_897H}) to compare with
the statistics from observational surveys since luminosity estimates
indicate that the stars are all giants. The number of J-type stars
($N_{{\rm J}}$) is then calculated according to Eq.~\ref{eq:counting},
along with the total number of C-stars ($N_{{\rm C}}$) to predict
the ratio $N_{\mathrm{J}}/N_{\mathrm{C}}$ giving the frequency of
J-stars among C-stars.

\section{Results}

\label{Results}The evolution of nova binaries that are of interest
in the re-accretion scenario for J-stars is presented in this section
followed by an estimate for the number fraction of such nova-polluted
C-stars expected to evolve as J-stars. 
\begin{table*}
\begin{centering}
\begin{tabular}{|c|c|c|c|c|c|c|c|}
\hline 
$M\mathrm{\mathrm{_{1}}}/M_{\odot}$ & $M\mathrm{_{2}}/M_{\odot}$ & $a/R_{\odot}$ & Case & $\mathrm{\mathrm{C/O}}$ & $\mathrm{\mathrm{^{12}C/^{13}C_{J}}}$ & $M\mathrm{_{J}}/M_{\odot}$ & $M\mathrm{_{Bol,J}}$\tabularnewline
\hline 
\hline 
$1.63$ & $1.20$ & $363.1$ & $\mathrm{COAL-A}$ & $1.03$ & $2.46$ & $1.21$ & $-3.48$\tabularnewline
\hline 
$2.15$ & $0.78$ & $631.0$ & $\mathrm{COAL-B}$ & $0.6$ & $9.00$ & $1.29$ & $-4.69$\tabularnewline
\hline 
$1.71$ & $0.94$ & $631.0$ & $\mathrm{CE-A}$ & $1.17$ & $2.09$ & $1.02$ & $-2.25$\tabularnewline
\hline 
$2.05$ & $1.39$ & $524.8$ & $\mathrm{CE-B}$ & $0.82$ & $3.16$ & $1.63$ & $-5.24$\tabularnewline
\hline 
\end{tabular}
\par\end{centering}

\caption{\label{tab:Post-nova-merger-cases} Initial binary parameters $M_{1},\, M_{2}$
and $a$ for typical post-nova merger progenitors with for $Z=0.008,\,\alpha_{\mathrm{CE}}=0.2$.
i) the post-nova merger emerges with surface $\mathrm{\mathrm{C/O}}$
as given in column 5. ii) The J-suffix denotes the instant at which
the merger has both $\mathrm{C/O}>1$~and~$\mathrm{\mathrm{^{12}C/^{13}C}<10}$
(J-star criteria) - thus $M_{{\rm J}}$ and $M_{{\rm Bol,J}}$ respectively
denote the mass and bolometric magnitudes (Eq.\ \ref{eq:mbol}) of
the post-nova merger when it classifies as a J-star.}
\end{table*}

\subsection{Evolution of WD-companions in nova binaries}

The orbital evolution of binary systems that emerge from a first CE
phase as a WD-MS binary depends on the mass ratio $q_{{\rm donor}}$
~(\citealt{1988A&A...191...57P}) given by,
\begin{equation}
q_{{\rm donor}}=\frac{M_{\mathrm{donor}}}{M_{\mathrm{WD}}}\,,
\end{equation}
where $M_{\mathrm{donor}}$ is the mass of the WD-companion. Assuming
conservative mass-transfer, the orbital separation decreases during
the nova phase if $q_{{\rm donor}}>1$, and $M_{{\rm donor}}$ decreases
until $q_{{\rm donor}}<1$. Thereafter, the orbit widens and the ultimate
fate of such a system is a detached state with $M_{\mathrm{donor}}$
being too low to evolve up to the giant branch within a Hubble time.
Hence such systems are not interesting as progenitors for nova-polluted
J-stars.

\subsubsection{Post-nova mergers as J-stars}

\label{sub:cases}The ultimate fate of systems which keep $q_{{\rm donor}}>1$
during the nova phase depends critically on the rate of mass-transfer,
$\dot{M}$. 
\begin{figure*}
\begin{singlespace}
\begin{centering}
\includegraphics[clip,angle=270,scale=0.6]{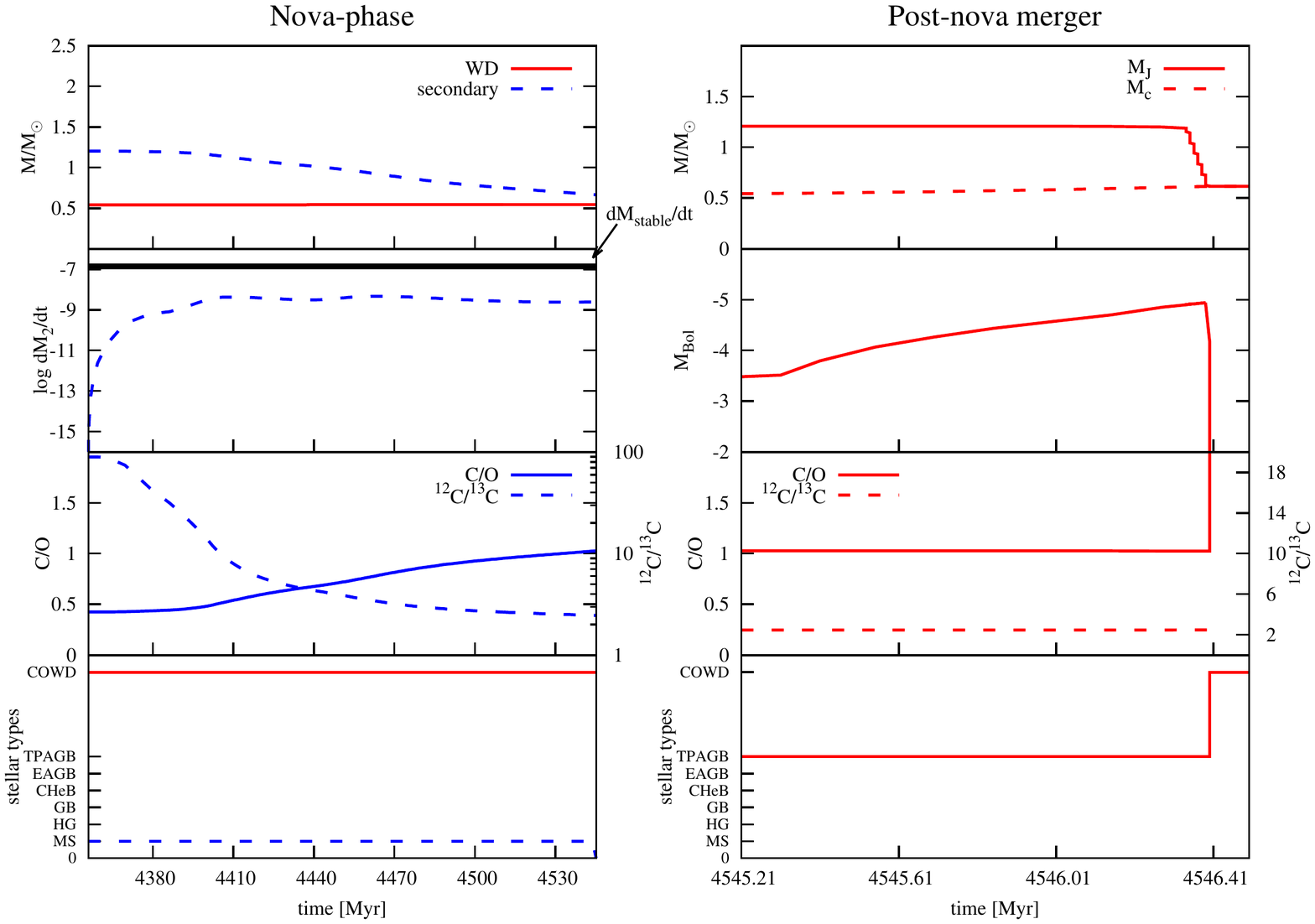}
\par\end{centering}

\begin{centering}
\textbf{A) COAL-A}
\par\end{centering}

\begin{centering}
\includegraphics[clip,angle=270,scale=0.6]{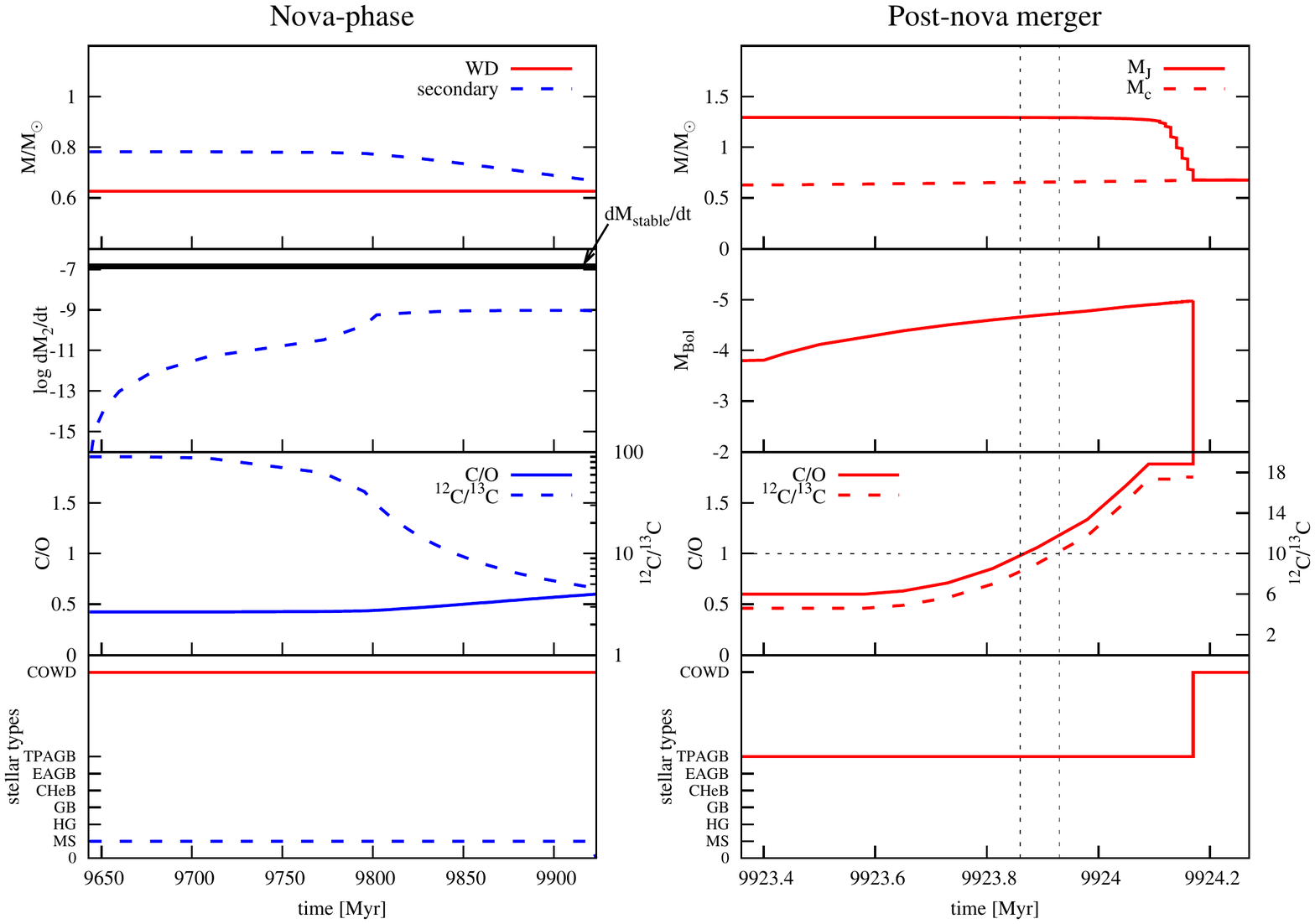}
\par\end{centering}

\begin{centering}
\textbf{B) COAL-B}
\par\end{centering}
\end{singlespace}

\caption{\label{fig:Typical-examples-of COAL}Evolution of COAL type systems
of Table~\ref{tab:Post-nova-merger-cases}. For each system i.e.
COAL-A (Fig. A) and COAL-B (Fig. B), the top and bottom panels on
the left show masses and evolutionary stages of both the primary (WD)
and the secondary stars during the nova phase when the mass-transfer
rate ($\frac{d\dot{M}_{2}}{dt}$) and surface chemistry ($\mathrm{C}/\mathrm{O}$,
$^{12}\mathrm{C}/^{13}\mathrm{C}$) for the donor (secondary) star
evolves with time as depicted in the second and third panels. The
right hand panels follow the post-nova merger properties i.e. mass,
core-mass, luminosity, surface chemistry and its stage of evolution
(EAGB/TPAGB). The COAL-A merger classifies as a J-star for the entire
duration of its AGB lifetime, as compared to the the COAL-B merger
that has $\mathrm{C}/\mathrm{O}>1$ during TDU on the TPAGB resulting
in a much shorter J-star phase marked by the vertical dotted lines.}
\end{figure*}
The systems in Table~\ref{tab:Post-nova-merger-cases} represent
typical cases in which the binary eventually merges into a giant single
star that satisfies the adopted criteria for J-stars (Section~\ref{sub:Population-Synthesis}).
Such post-nova mergers classify into two distinct binary evolutionary
channels depending on the evolution of mass-transfer during the nova
phase:
\begin{enumerate}
\item \textbf{Coalescence} (\textbf{COAL}): If the nova phase continues
with $\dot{M}<10^{-7}\, M_{\odot}\mathrm{yr^{-1}}$ until the WD companion
(donor) becomes a low-mass ($M_{\mathrm{donor}}\simeq0.7\, M_{\odot}$)
MS star, mass-transfer to the WD becomes dynamically unstable as the
donor (MS) star becomes deeply convective and $q_{{\rm donor}}\mathrm{\,(>1)}$
also exceeds the critical mass-ratio $q\mathrm{_{crit}}$ for stable
RLOF (\citealt{2002MNRAS_329_897H}). All the material is accreted
and swells up to form a giant envelope around the WD which becomes
the core of a merged giant (TPAGB) star.\\
If re-accretion of nova-ejecta pollutes the WD-companion such that
its surface $\mathrm{C}/\mathrm{O}>1$ and $^{12}\mathrm{C}/^{13}\mathrm{C}<10$,
the coalesced (AGB) star classifies as a J-star. Such a typical system
is labeled as COAL-A in Table~\ref{tab:Post-nova-merger-cases} in
which the merged star is a $1.2\, M_{\odot}$ TPAGB star with a CO
core mass $\sim0.6\, M_{\odot}$ and hence has a lower luminosity
$L\sim10^{3}\, L\mbox{\ensuremath{_{\odot}}}$\ $(\Rightarrow M_{{\rm Bol}}>-4)$
compared to typical AGB C(N)-stars with $L\sim10^{4}\, L\mbox{\ensuremath{_{\odot}}}$\ $(\Rightarrow M_{{\rm Bol}}<-4)$
and $M\gtrsim2\, M_{\odot}$. . However, following the nova-phase,
if the WD-companion has surface $\mathrm{C}/\mathrm{O}<1$ but $^{12}\mathrm{C}/^{13}\mathrm{C}<10$,
the merger can subsequently evolve on the AGB with $\mathrm{C}/\mathrm{O}>1$
because of TDU until its $^{12}\mathrm{C}/^{13}\mathrm{C}>10$, and
classify as a J-star. A typical example is labeled as COAL-B in Table~\ref{tab:Post-nova-merger-cases}
with a higher luminosity ($M_{{\rm Bol}}<-4$) as compared to the
COAL-A system depending on their evolutionary phases on the AGB when
they classify as J-stars.\textbf{ }Fig.~\ref{fig:Typical-examples-of COAL}
illustrates the differences in evolution of these two typical COAL
systems through the nova-phase followed by the AGB phase of the nova-polluted
merger that evolves into a J-star.
\item \textbf{Common envelope }(\textbf{CE}):\textbf{ }For suitable initial
parameters ($M\mathrm{_{1}}$, $M\mathrm{_{2}}$, $a$), e.g. systems
CE-A/B of Table~\ref{tab:Post-nova-merger-cases}, RLOF proceeds
at the thermal rate of the WD-companion when it evolves as a Hertzsprung
Gap (HG) star with a radiative or thin convective envelope. When the
thermal rate is less than $10^{-7}\, M_{\odot}\mathrm{\, yr^{-1}}$,
the nova phase continues until the WD-companion becomes a giant (GB)
when mass-transfer is dynamically unstable and leads to a (second)
CE phase. Consequently the system merges and the CO-WD becomes the
core surrounded by a He-shell (core of the giant donor) resembling
an EAGB star whose surface abundances are determined primarily by
the donor star which suffers from nova-pollution. A typical case is
labelled as CE-A in Table~\ref{tab:Post-nova-merger-cases} with
the merged EAGB star having both $\mathrm{C}/\mathrm{O}>1$ and $^{12}\mathrm{C}/^{13}\mathrm{C}<10$
.\textbf{}\\
If the thermal rate exceeds $10^{-7}\, M{}_{\odot}\,\mathrm{yr^{-1}}$
during the RLOF-phase, steady H-burning on the surface of the accreting
WD ends the nova phase. Thereafter, the orbit continues to shrink
leading to runaway mass transfer until the material accreted onto
the WD swells up to form a giant star and the system also merges in
a (second) CE-phase.Such an example is the CE-B case of Table~\ref{tab:Post-nova-merger-cases}
for which the CE merger outcome ascends the AGB with $^{12}\mathrm{C}/^{13}\mathrm{C}<10$
owing to previous nova-pollution until it becomes a C-star when $\mathrm{C}/\mathrm{O}>1$
because of TDU. Thus it classifies as a J-star only for a brief phase
($\sim10^{5}\,\mathrm{yr}$) of its TPAGB until $^{12}\mathrm{C}/^{13}\mathrm{C}>10$
because $^{12}\mathrm{C}$ is brought to the surface by TDU.\textbf{
}The evolution of this system is shown in Fig.~\ref{fig:Typical-examples-of CE}
for the nova phase preceding the CE, and the AGB phase of the merged
star that becomes a J-star on the EAGB for the CE-A system but requires
TDU on the TPAGB for the CE-B system.\textbf{ }
\begin{figure*}
\begin{singlespace}
\begin{centering}
\includegraphics[clip,angle=270,scale=0.65]{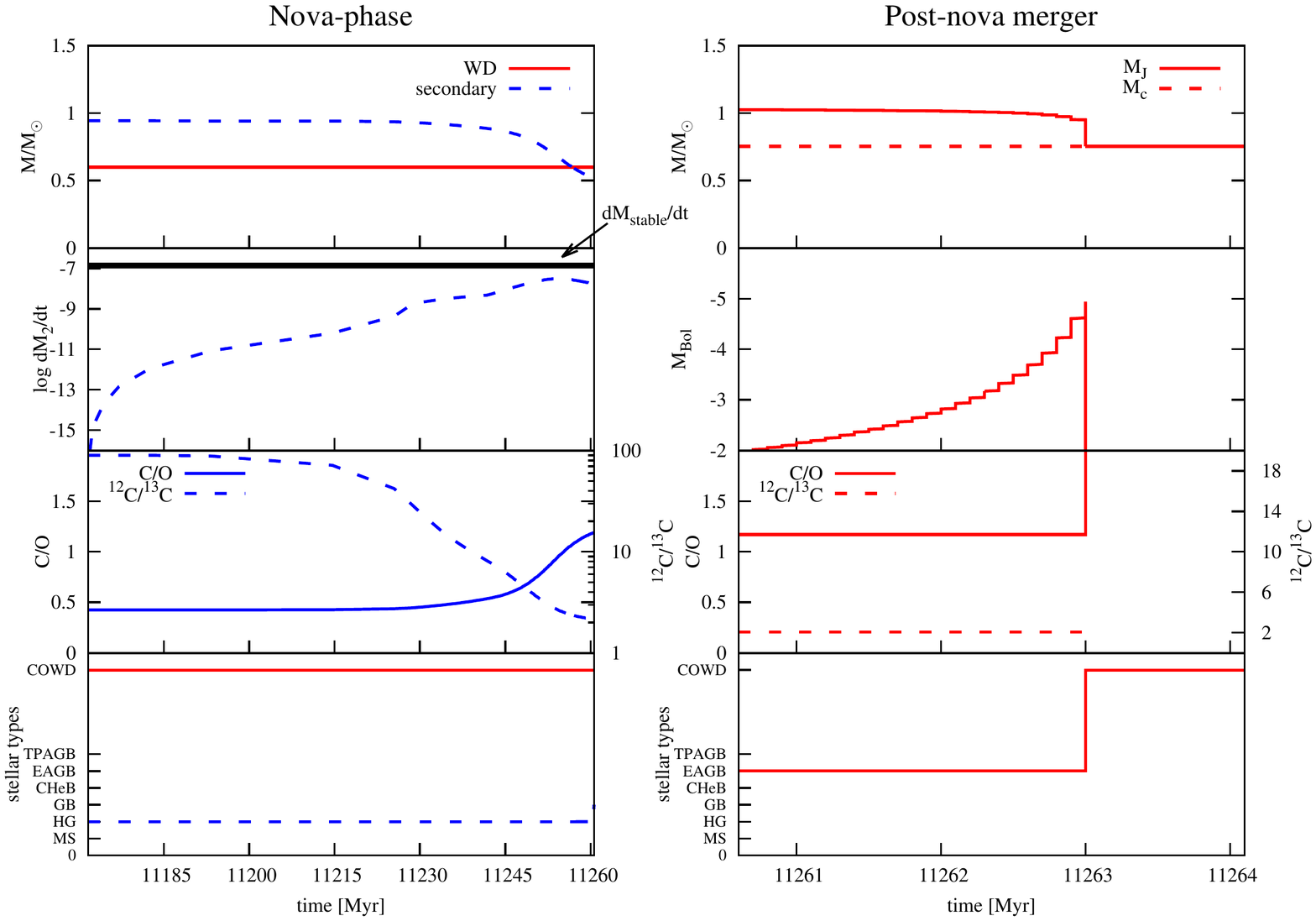}
\par\end{centering}

\begin{centering}
\textbf{A) CE-A}
\par\end{centering}

\begin{centering}
\includegraphics[clip,angle=270,scale=0.65]{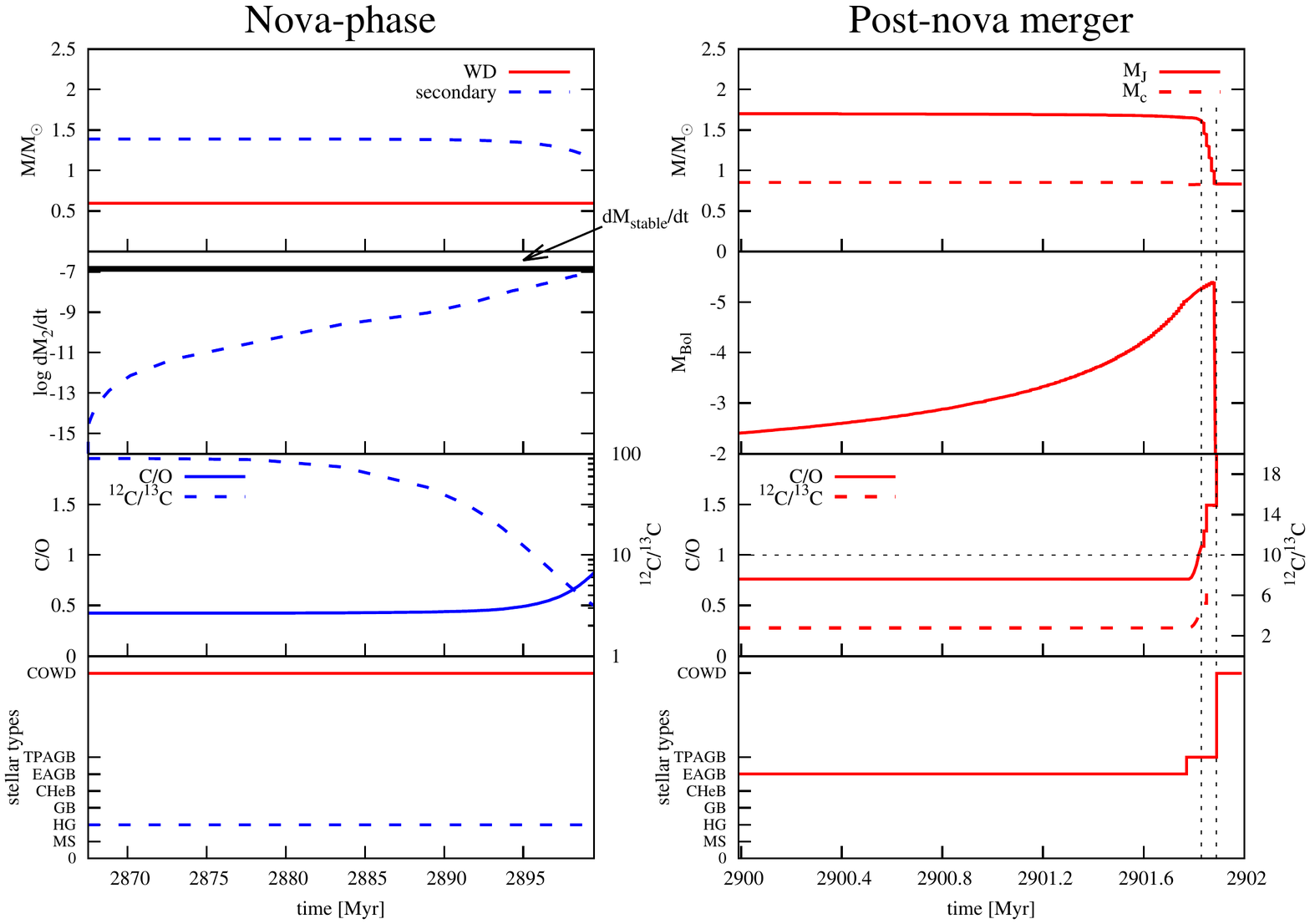}
\par\end{centering}

\begin{centering}
\textbf{B) CE-B}
\par\end{centering}
\end{singlespace}

\caption{\label{fig:Typical-examples-of CE}Evolution of CE type systems of
Table~\ref{tab:Post-nova-merger-cases}. Refer to Fig~\ref{fig:Typical-examples-of COAL}
caption for description of evolutionary stages shown. The CE-A merger
exhibits properties of a J-star on the EAGB unlike the COAL-B merger
which has $\mathrm{C}/\mathrm{O}>1$ only during TDU and hence has
a short phase on the TPAGB (marked by the dotted lines) when it classifies
as a J-star.}
\end{figure*}

\end{enumerate}
All the above examples of nova-binaries considering pollution of the
WD-companion stars present interesting outcomes that can classify
as J-stars with their evolutionary status determined by the preceding
nova-phase of the progenitor systems. Accordingly, they vary in their
properties (viz. masses, luminosities, chemistry) depending on whether
the WD-companion re-accretes enough nova ejecta to evolve with $\mathrm{C}/\mathrm{O}>1$,
along with $^{12}\mathrm{C}/^{13}\mathrm{C}<10$. 

Thus, the nova re-accretion scenario predicts mergers that evolve
as J-stars with very low $^{12}\mathrm{C}/^{13}\mathrm{C}\,(<4)$
ratios and luminosities $L\sim10^{3}\, L\mbox{\ensuremath{_{\odot}}}$\ $(\Rightarrow M_{{\rm Bol}}>-4)$
unlike typical AGB C-stars undergoing HBB with low $^{12}\mathrm{C}/^{13}\mathrm{C}$.
Post-nova merger systems like COAL-A and CE-A are C-rich from the
re-accretion of nova-ejecta and thus classify as J-stars even without
TDU. On the other hand, cases COAL-B and CE-B which emerge as post-nova
mergers necessarily require TDU to have $\mathrm{C/O>1}$ and thus
behave as more luminous J-stars on the TPAGB.

\subsubsection{Initial binary parameters for post-nova mergers}

\begin{figure}
\begin{centering}
\includegraphics[angle=270,scale=0.35]{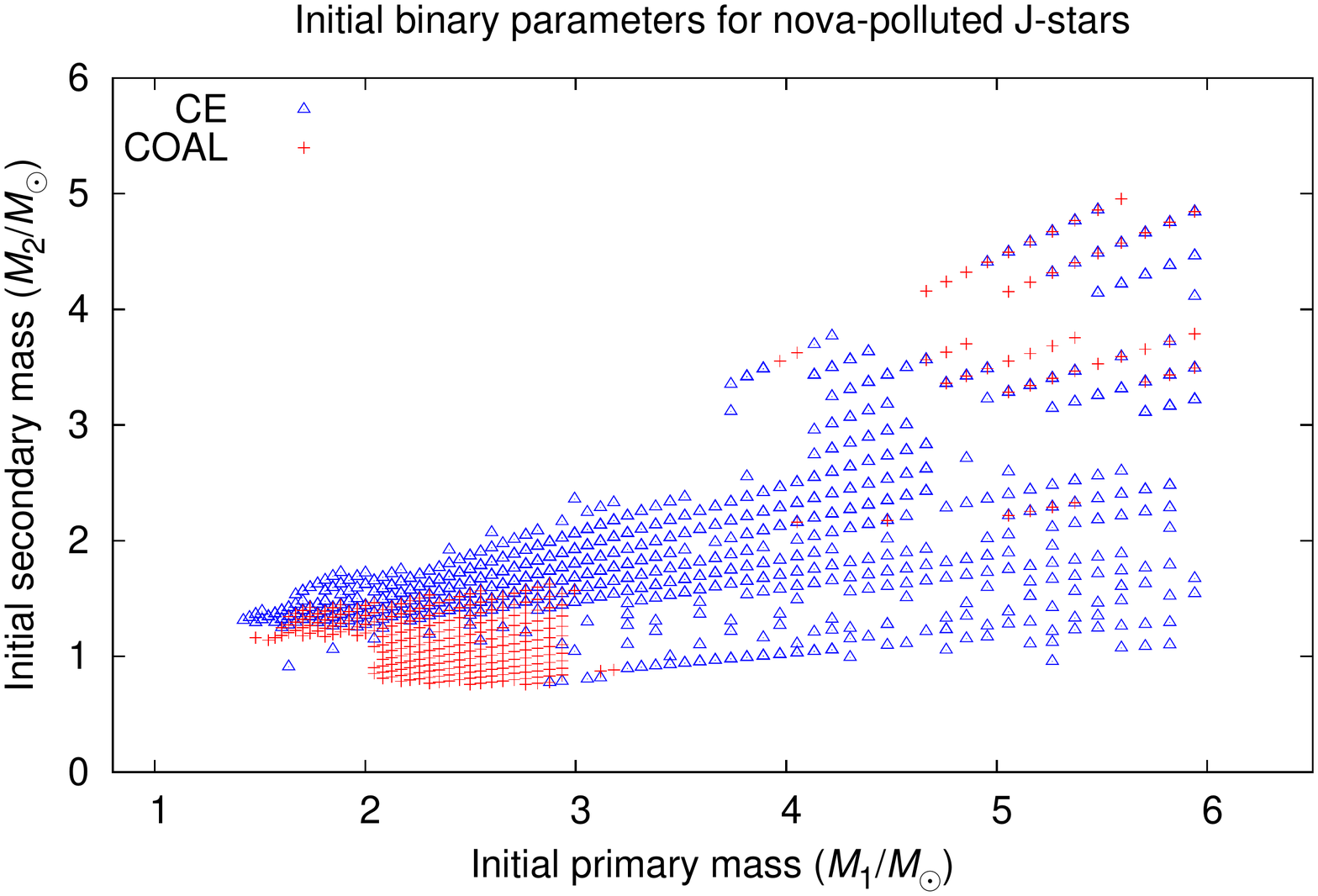}
\par\end{centering}

\begin{centering}
\includegraphics[angle=270,scale=0.35]{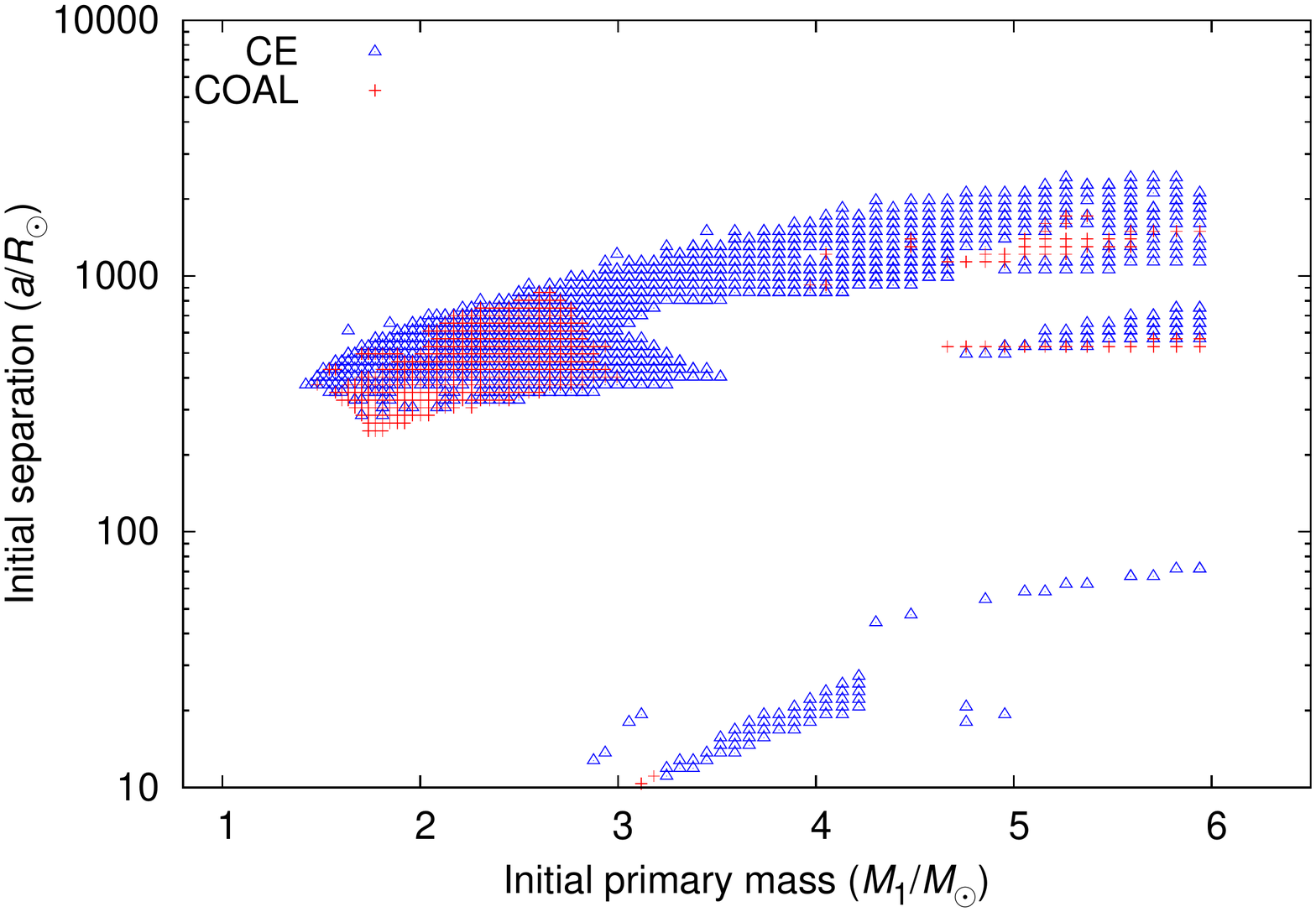}
\par\end{centering}

\caption{\label{fig:Range-of-initial-parameters}Range of initial primary mass
$M_{1},$ initial secondary mass $M_{2}$ and initial separation $a$
for nova-merger cases (COAL/CE) that form J-stars.}
\end{figure}

Fig.~\ref{fig:Range-of-initial-parameters} shows the range of initial
binary parameters ($M_{{\rm 1}}\mathrm{\mbox{ vs }}M_{{\rm 2}}$,
$M_{{\rm 1}}\mbox{ vs }a$) for the post-nova merger s that evolve
to become J-stars.
The COAL mergers that classify as J-stars mostly originate from binary
systems with WD masses close to $0.6\, M_{\odot}$ for $M_{1}<3\, M_{\odot}$.
CE mergers mostly originate from systems with $M_{2}>1\, M{}_{\odot}$
which involve thermal-timescale mass-transfer (mostly on the HG) during
RLOF. 
\begin{table*}[t]
\renewcommand{\multirowsetup}{\centering}

\begin{centering}
\begin{tabular}{|c|c|c|c|c|c|}
\hline 
\multirow{2}{*}{$Z$} & \multirow{2}{*}{$N_{\mathrm{J}}/N_{\mathrm{C}}(\%)$ from single (HBB) stars (I)} & \multicolumn{2}{c|}{$N_{\mathrm{J}}/N_{\mathrm{C}}(\%)$ from binaries (II)} & \multicolumn{2}{c|}{ $N_{\mathrm{J}}/N_{\mathrm{C}}(\%)$ (III)}\tabularnewline
\cline{3-6} 
 &  & post-nova mergers (II a) & HBB (II b) & model (III a) & obs. (III b)\tabularnewline
\hline 
\hline 
\multirow{2}{2cm}{0.008} & \multirow{2}{1.5cm}{0.71} & $3.0$ & $0.27$ & $1.95$ & \multirow{2}{1cm}{10}\tabularnewline
\cline{3-5} 
 &  & $1.07^{*}$ & $4.75^{*}$ & $4.45$ & \tabularnewline
\hline 
\multirow{2}{2cm}{0.02} & \multirow{2}{1.5cm}{0.00} & $0.3$ & $0.06$ & $0.36$ & \multirow{2}{1cm}{15}\tabularnewline
\cline{3-5} 
 &  & $0.01^{*}$ & $3.43^{*}$ & $3.44$ & \tabularnewline
\hline 
\end{tabular}
\par\end{centering}

\caption{\label{tab:-ratio-estimated}$N_{{\rm J}}/N_{{\rm C}}$ ratios for
single (Col I) and binary (Col II) stars. i) The ratios in IIa and
IIb marked by $*$ are calculated by including GB-MS (Case-B RLOF)
mergers which have a longer HBB-phase compared to single AGB stars
of similar total mass, and hence increase the $N_{\mathrm{C}}$ and
\foreignlanguage{english}{$N_{\mathrm{J}}$} for binaries from HBB
stars. ii) The model (III a) $N_{\mathrm{J}}/N_{\mathrm{C}}$ is calculated
by adding the contributions to $N_{\mathrm{J}}\mbox{ and }N_{\mathrm{C}}$
from both single (I) and binary stars (II a \& b), assuming a constant
binary fraction of $0.5$. iii) Observed statistics (III b) are from
\citet{2003MNRAS.341..534M} and \citet{2000ApJ...536..438A} for
the LMC and our Galaxy respectively.}
\end{table*}

\subsection{\label{par:Role-of-binary} J-stars from Case-B RLOF mergers }

\begin{figure}
\begin{centering}
\includegraphics[scale=0.4]{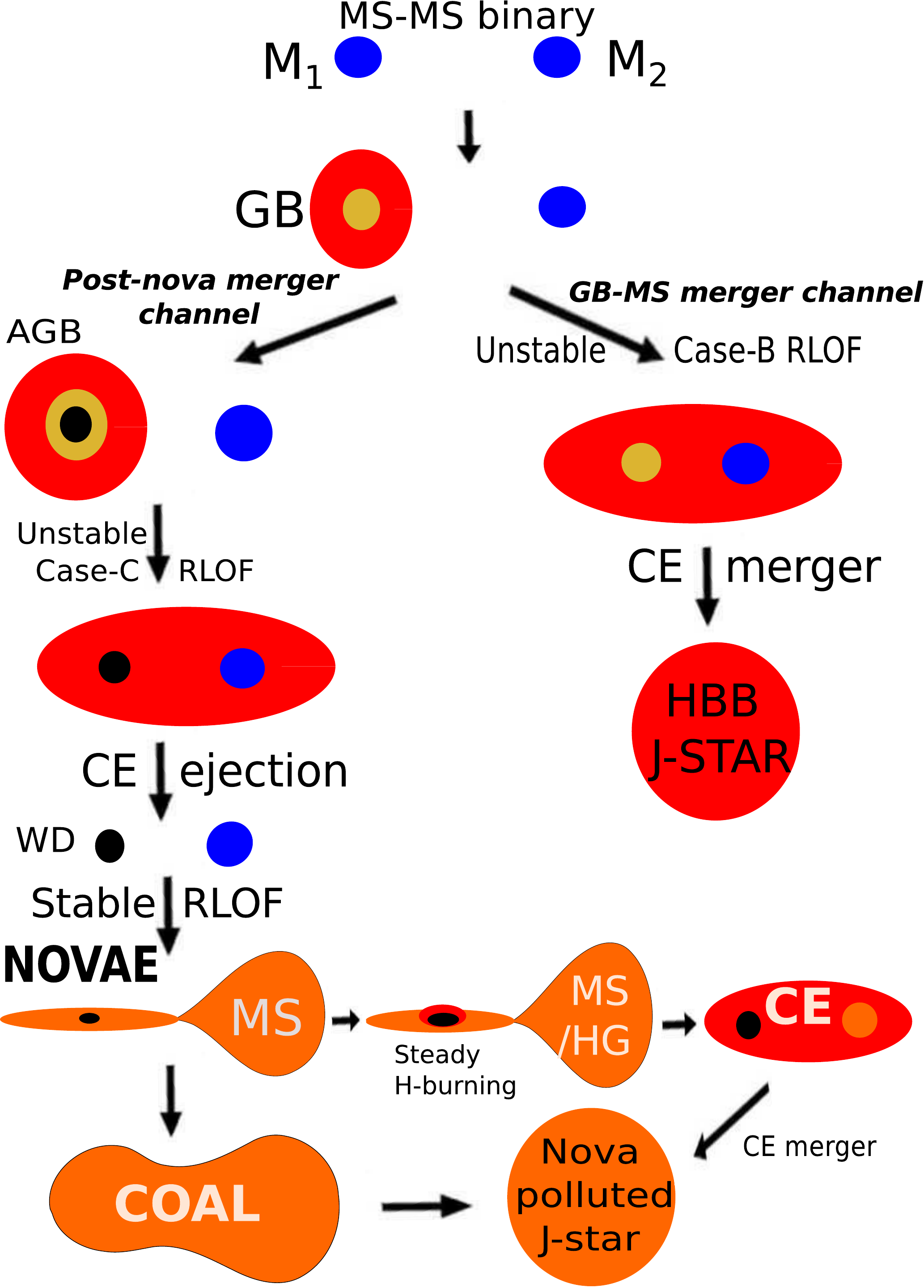}\caption{\label{fig:Binary-evolutionary-channels}Binary evolutionary channels
which lead to mergers with properties of J-stars from re-accretion
of nova-ejecta followed by coalescence (COAL) and common-envelope
(CE) mergers or HBB in AGB-phases of GB-MS (Case-B RLOF) mergers.
The different stellar evolutionary phases are marked (with colours)
as - blue: MS star, red: giant-envelope/ CE, golden: He-core/WD, black:
CO-core/WD, orange: nova-polluted star (MS/HG).}

\par\end{centering}

\end{figure}

Unlike nova progenitor systems that emerge as a WD-MS binary following
a CE phase, depending on their initial separation binaries also merge
in the CE phase because of unstable mass transfer from a primary star
on the GB to a MS secondary (Case-B RLOF). This is illustrated in
Fig.~\ref{fig:Binary-evolutionary-channels} which distinguishes
the evolutionary channel for these systems from the post-nova mergers
(COAL/CE) as described in Section \ \ref{sub:cases}. The merged
(GB) star has a lower helium core mass than expected for a single
star of the same total mass at the beginning of the GB. The typical
evolution of such an object on the TPAGB is shown in Fig.~\ref{fig:TPAGB-phase}
(cf. Appendix~B), along with a single star model of similar total
mass, which illustrates the difference in evolutionary time-scales
for the two cases. As a result, they contribute significantly to the
number of stars that classify as C and J-stars in our population synthesis
calculations as described next.

\subsection{\label{sub:Population-Synthesis-1}Population synthesis }

With the method outlined in Section~\ref{sub:Population-Synthesis},
a binary grid of resolution $100\times50\times100$ in $M_{1}-M_{2}-a$
space is evolved for $Z=0.008$\ and\ $0.02$,$\,\alpha_{\mathrm{CE}}=0.2$
and $f_{\mathrm{bin}}=0.5$ to estimate the total number of J-stars
expected from the instances of nova-pollution as described in Section~\ref{sub:cases},
as well as from HBB in both single and binary AGB stars.

\subsubsection{Observational selection effects}

Our population synthesis is performed considering possible selection
effects in observations from existing surveys for J-stars. For instance
in the LMC survey of \citet{2003MNRAS.341..534M}, the J, H, K magnitudes
for all J-stars lead to bolometric magnitudes $M\mathrm{_{Bol}}<-2.8$
derived using the relation (as adopted in \citealt{2003MNRAS.341.1290H}), 

\begin{equation}
M_{\mathrm{Bol}}=K_{0}-dm+0.69+2.65(J-K)_{0}-0.67(J-K)_{0}^{2}\,,\label{eq:mbol}
\end{equation}
with a distance modulus $dm=18.45$ for the LMC (\citealt{1997macl.book.....W}),
a K-band absorption $A\mathrm{_{K}=0.02}$ (\citealt{1983ApJ...272...99W})
along with mean reddening $E(J-K)=0.07$ (\citealt{1996AJ....112.2607C})
to find the de-redenned colours denoted by $K_{0}=K-A_{{\rm K}}$
and $(J-K)_{0}=(J-K)-E(J-K)$. We use this as a low-luminosity cut-off
in our estimates for the number of C ($N_{{\rm C}}$) and J ($N_{{\rm J}}$)
stars.

\subsubsection{\label{sub:-ratio}$N_{{\rm J}}/N_{{\rm C}}$ ratio}

The number of stars of interest (i.e. $N_{\mathrm{J}},\, N_{\mathrm{C}}$)
is calculated as described in Section~\ref{sub:Population-Synthesis}
for all systems both in our single and binary grids for metallicities
$Z=0.008$ (LMC) and $0.02$ (solar). The resulting $N_{{\rm J}}/N_{{\rm C}}$
ratio are presented in Table~\ref{tab:-ratio-estimated}, with the
binary J-stars separated (in columns) according to their evolutionary
origin i.e. post-nova merger or HBB. Furthermore, the difference in
the values of the ratios between the upper and lower rows in each
of these two columns occurs from counting Case-B RLOF mergers (as
described in Section~\ref{par:Role-of-binary}) which have considerably
longer HBB-AGB phases compared to single AGB stars of same total mass
(cf. Appendix\ B) and classify as J-stars. However, the total $N_{{\rm J}}/N_{{\rm C}}$
ratio considering all single (HBB) and binary (merger) J-stars is
still below the observed fraction ($10-15\%)$ for both $Z=0.008$\ and\ $0.02$.

\subsubsection{Luminosity function for LMC J-stars}

\begin{figure}
\begin{centering}
\includegraphics[angle=270,scale=0.35]{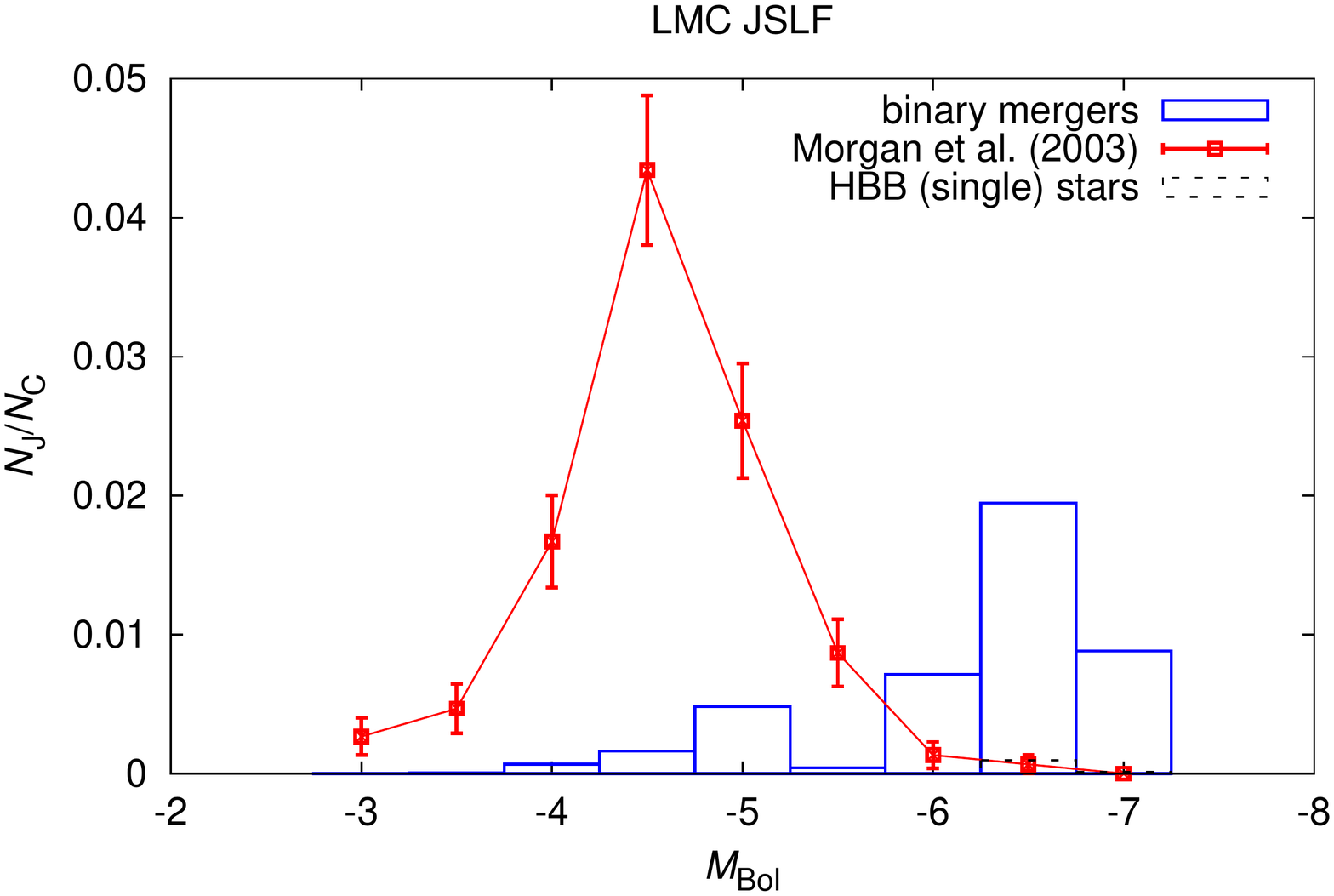}
\par\end{centering}

\caption{\label{fig:Luminosity-function}Luminosity function of LMC J-stars
(JSLF). The histograms binned in $M\mathrm{_{Bol}}$ intervals of
0.5 mag show expected number fraction ($N_{{\rm J}}$/$N_{C}$) of
J-stars among C-stars for HBB (single) stars and binary mergers (post-nova
or Case-B RLOF) - compared with the JSLF constructed for the LMC J-stars
of \citealt{2003MNRAS.341..534M} with Poisson $\sqrt{n}$ error bars
where $n$ is the number of stars in a particular $M\mathrm{_{Bol}}$
bin.}
\end{figure}

The J-star luminosity function (JSLF) is constructed with the $N_{{\rm J}}/N_{{\rm C}}$
ratio binned in $M\mathrm{_{Bol}}$ intervals of 0.5 mag. (Fig.~\ref{fig:Luminosity-function}).
In order to compare with observations, Eq.~\ref{eq:mbol} is used
to derive the bolometric magnitudes of all the 156 LMC J-stars of
\citet{2003MNRAS.341..534M}. 

Fig.~\ref{fig:Luminosity-function} shows that the post-nova merger
channel produces J-stars that are dimmer (with $M\mathrm{_{Bol}}>-5$)
but much rarer compared to the HBB stars (with $M\mathrm{_{Bol}}<-6$).
In contrast, the distribution constructed from the observed sample
of \citet{2003MNRAS.341..534M} has the majority of J-stars dimmer
than $M_{\mathrm{Bol}}=-6$. 
\begin{figure}
\begin{centering}
\includegraphics[angle=270,scale=0.35]{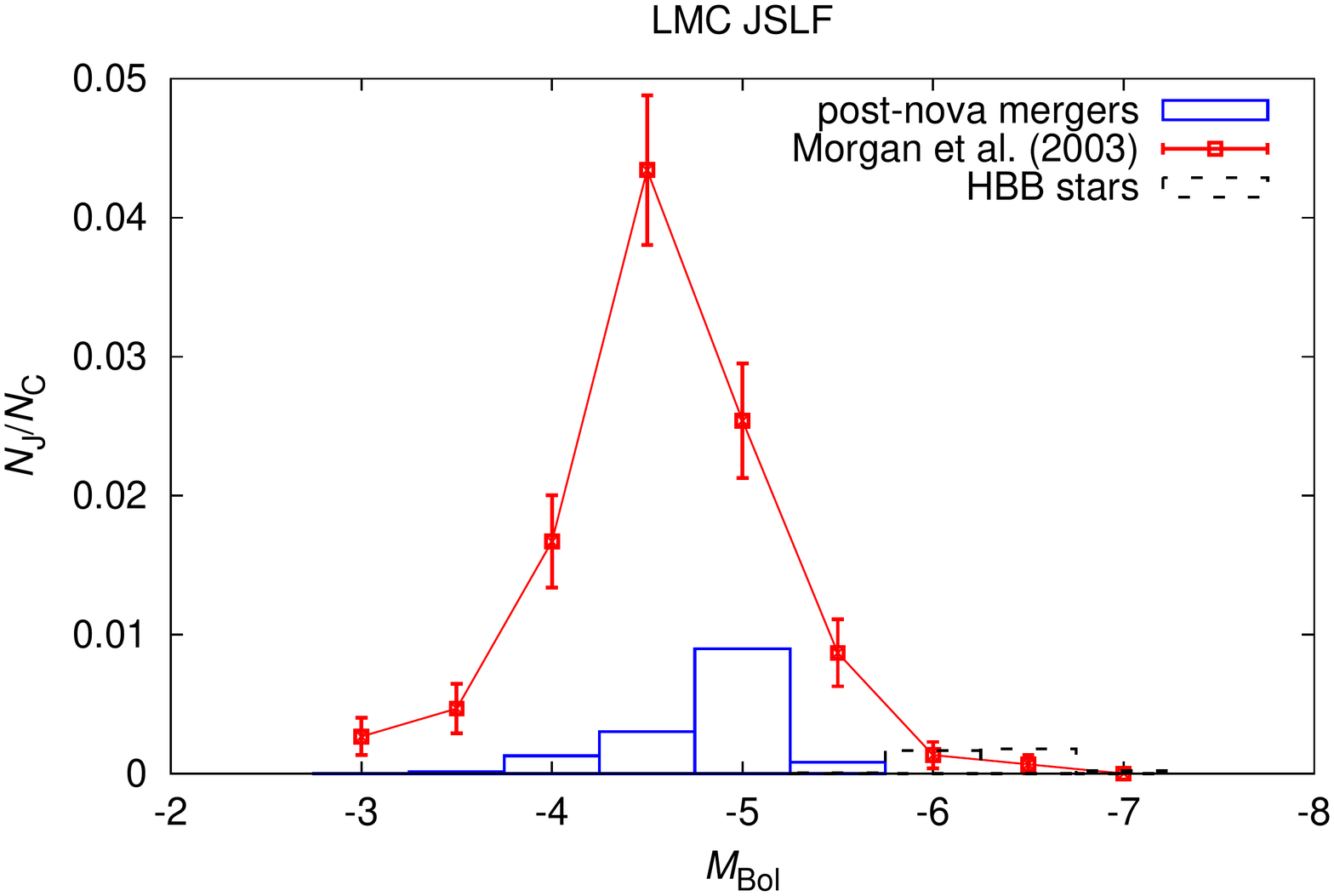}
\par\end{centering}

\caption{\label{fig:LMC-JSLF-cosidering} Luminosity function of LMC J-stars
(JSLF). Number fraction ($N_{{\rm J}}$/$N_{C}$) of J-stars among
C-stars for post-nova mergers and HBB-stars excluding Case-B RLOF
binary mergers is shown with observed J-stars from the sample of \citealt{2003MNRAS.341..534M}
binned in $M\mathrm{_{Bol}}$ intervals of 0.5 mag. The total $N_{{\rm J}}$/$N_{C}$
ratio is $\sim2\%$ for the nova-polluted mergers - lower than the
total $N_{{\rm J}}$/$N_{C}$ in Fig.\ \ref{fig:Luminosity-function}
($\sim5\%)$ and the observed fraction ($\sim10\%$) of the stars
in the sample of \citealt{2003MNRAS.341..534M}.}
\end{figure}
However, most of the J-stars with $M\mathrm{_{Bol}}<-6$ result from
Case-B RLOF whose post-merger evolution on the (TP)AGB is different
compared to single (AGB) stars of similar masses (cf. Fig.~\ref{fig:TPAGB-phase}).
The JSLF for the post-nova mergers is shown separately in Fig.~\ref{fig:LMC-JSLF-cosidering}
along with that of single and binary HBB-AGB stars excluding the Case-B
RLOF mergers. It has closer resemblance to the JSLF of the stars from
\citet{2003MNRAS.341..534M}, with a higher fraction of low-luminosity
J-stars as observed in the \citealt{2003MNRAS.341..534M} survey,
though the predicted $N_{{\rm J}}/N_{{\rm C}}$ is still lower and
well below the observed fraction for the dimmer ($M\mathrm{_{Bol}}>-6$)
J-stars in the LMC.

\subsubsection{Number distributions}

\begin{figure}
\begin{centering}
\includegraphics[angle=270,scale=0.35]{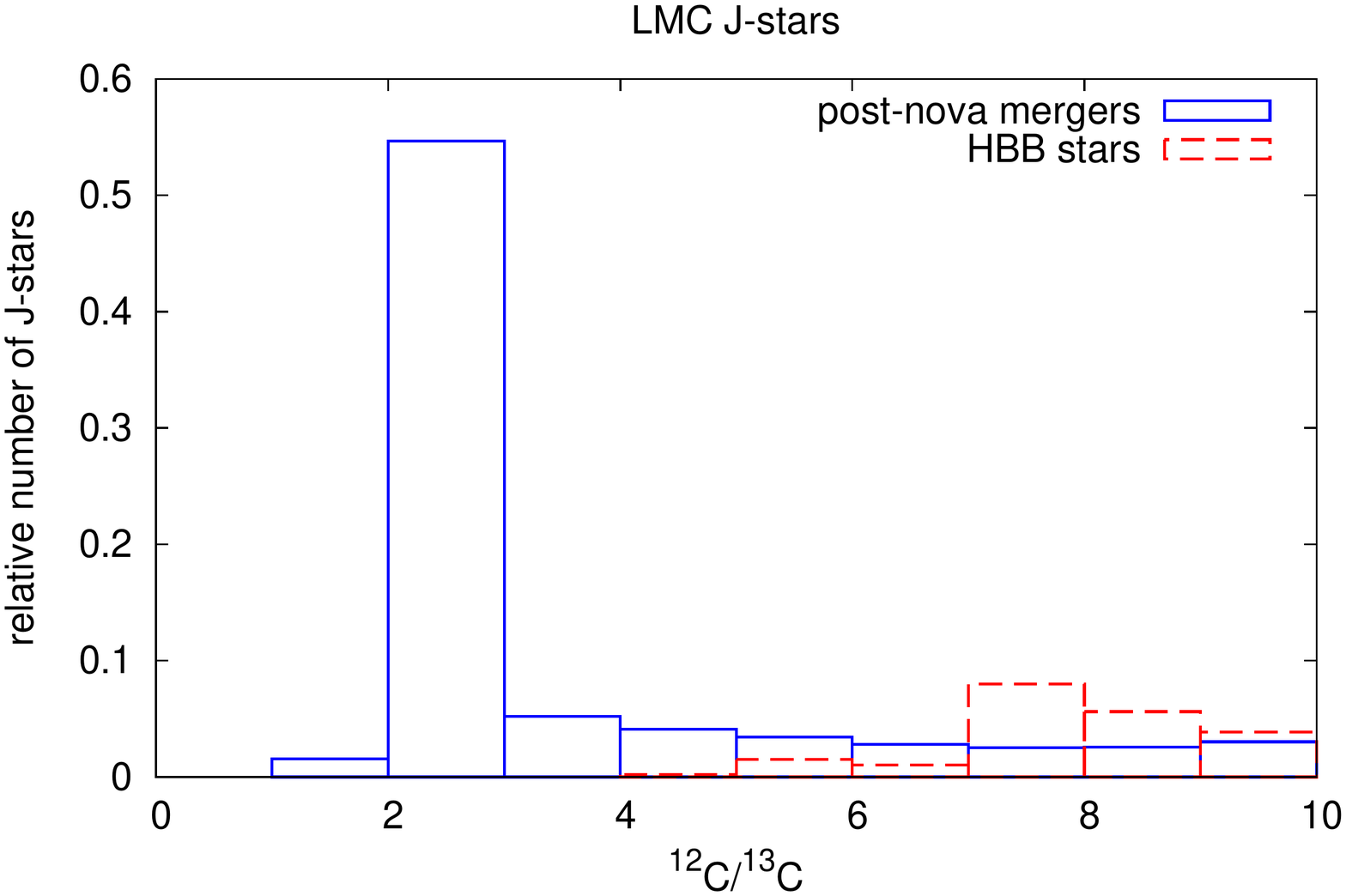}
\par\end{centering}

\begin{centering}
\includegraphics[angle=270,scale=0.35]{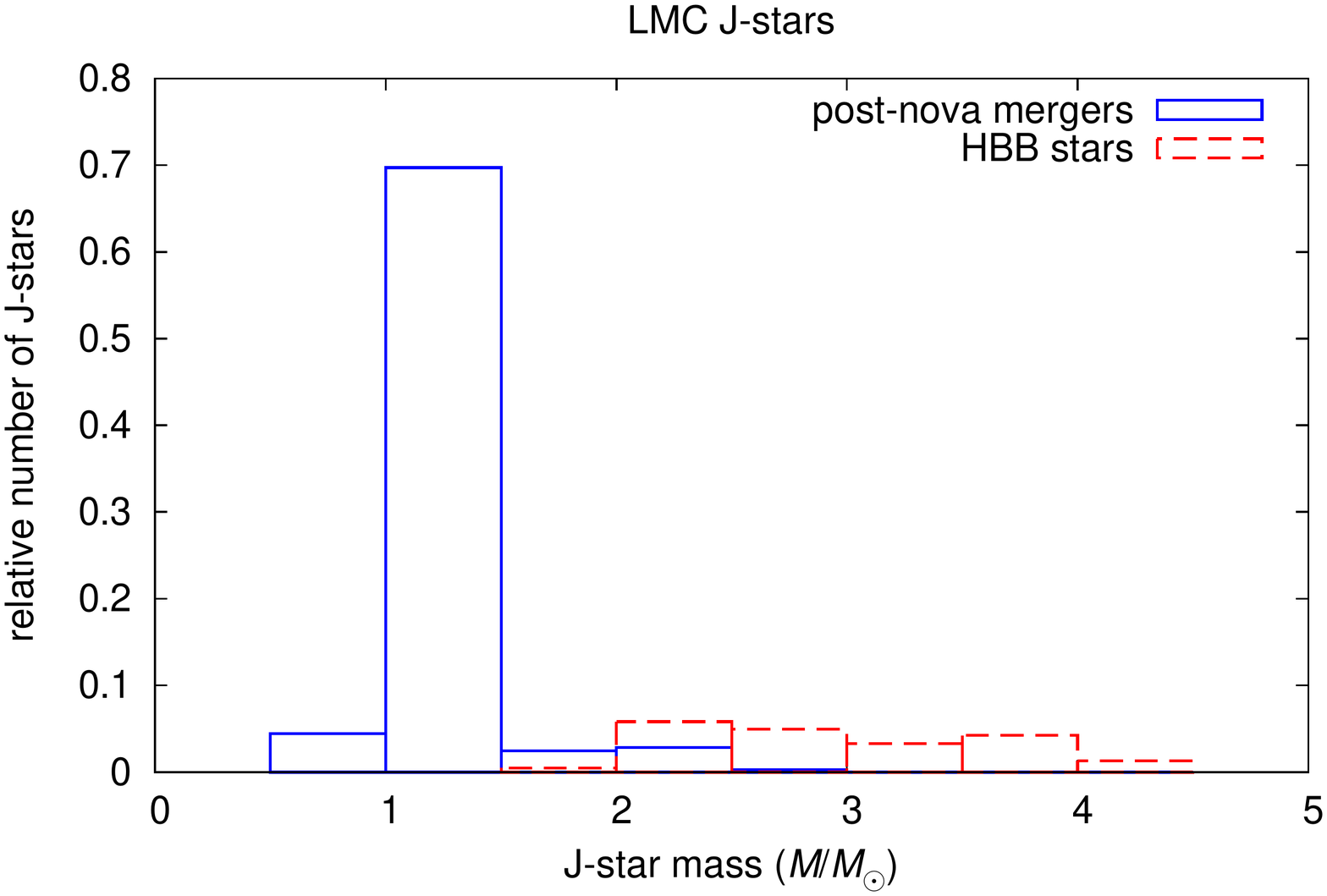}
\par\end{centering}

\caption{\label{fig:Number-distributions-12C/13C}Distributions of $\mathrm{^{12}C/^{13}C}$
(upper panel) and masses (lower panel) for J-stars including post-nova
mergers and HBB-stars but excluding Case-B RLOF binary mergers for
$Z=0.008$. The relative number of J-stars is the ratio of number
of J-stars in each bin to the total number of J-stars. }
\end{figure}

The number distributions for masses ($M$) and $^{12}\mathrm{C}/^{13}\mathrm{C}$
ratios of LMC ($Z=0.008$) J-stars are shown in Fig.~\ref{fig:Number-distributions-12C/13C},
excluding the HBB Case-B RLOF mergers (as described in Section~\ref{par:Role-of-binary}).
The post-nova mergers have $\mathrm{^{12}C/^{13}C}$ ratios lower
than $4$, and low masses ($<2\, M_{\odot}$) unlike the HBB stars,
since majority of the post-nova mergers owe their C-enhancement to
the re-accretion of C-rich nova ejecta from a low-mass ($\sim0.6\, M_{\odot}$)
CO-WD and can evolve as J-stars without TDU on the AGB. The $\mathrm{^{12}C/^{13}C}$
ratios remain low on the AGB for such systems e.g. COAL-A (cf. Table~\ref{tab:Post-nova-merger-cases})
whereas for HBB-stars ($>2\, M_{\odot}$) the $\mathrm{^{12}C/^{13}C}$
ratios increase during TDU when they become C(J)-stars.

\section{Discussion}

The binary scenario for the origin of J-stars explored in this work
illustrates the need for detailed evolutionary studies in several
aspects of both single and binary star evolution. The nucleosynthesis
associated both with nova explosions for low-mass CO-WDs as well the
treatment of HBB during AGB evolution, particularly in context of
binary mergers, require detailed models that account for the evolutionary
history of the associated binary system. Particularly, detailed evolutionary
calculations of RG-MS (Case-B RLOF) mergers are required to determine
their evolutionary time-scales and nucleosynthesis (especially the
phase with HBB) on the AGB.

In the context of post-nova mergers that can exhibit properties associated
with J-stars, other chemical peculiarities e.g. the O-isotopic ratios,
N-enhancements, Li-enrichment, s-process abundances etc., need to
be investigated and compared with observations (e.g. \citealt{2013ApJ...768L..11H}).
To quantitatively predict these isotopic ratios for the J-stars expected
from the nova-pollution scenario, a detailed low-mass CO-WD nova model
is required providing ejecta composition for initial ${\rm C}:{\rm O}$
ratio different from a uniform one (50:50) as used in existing models
of CO-WD explosions. Also, chemical pollution of the secondary (donor)
star during the nova phase with carbon rich ejecta should influence
the subsequent nova explosions when RLOF from the nova-polluted donor
transfers some of the material from its surface back to the WD. However,
investigating such an effect requires nova models with varying accretion
compositions which is beyond the scope of the present study. 

The systematic uncertainties associated with our prediction for the
$N_{\mathrm{J}}/N_{\mathrm{C}}$ ratio for the post-nova merger channel
(without HBB stars) include our choice of the CE ejection efficiency
parameter $\mathrm{\alpha_{CE}}$. The post-CE separations for the
WD-MS nova progenitor systems increase with $\mathrm{\alpha_{CE}}$
which prevents RLOF from the WD-companion during its MS/HG phase of
evolution and a subsequent nova phase for the binary. We adopt a constant
$\mathrm{\alpha_{CE}}=0.2$ based on our estimate for the Galactic
nova rate (refer Appendix~\ref{sec:Constraining-CE-ejection}) which
is constrained by observations \citep{2002AIPC..637..462S}. With
regard to other assumptions in our modelling of nova systems, the
use of a nearly constant WD mass in the nova phase is justifiable
within the usual range of classical novae mass transfer rates ($\dot{M}\sim10^{-10}-10^{-8}\, M_{\odot}\mathrm{\, yr^{-1}}$)
based on detailed models which investigate the ratio of mass ejected
to that accreted in outbursts for the entire range of $M_{\mathrm{WD}}$
\citep{2005ApJ...623..398Y}. We also use a constant binary fraction
$f_{{\rm bin}}=0.5$ in our population synthesis since the post-nova
mergers that classify as J-stars evolve from low-mass WD-MS systems
and in view of existing observations estimating the binary fraction
for this range ($\sim1\, M_{\odot}$) of stellar masses (\citealt{2013arXiv1303.3028D}).

\section{Conclusions}

This work investigates the effect of chemical pollution of WD companions
in nova binaries because of re-accretion of nova ejecta and their
subsequent evolution with observed properties of J-stars. The re-accretion
model (Section~\ref{sub:The-Nova-Re-accretion}) is applied to a
population of binary systems using the population nucleosynthesis
code \textit{binary\_c/nucsyn} to predict the properties of post-nova
systems that evolve as giant stars having both $\mathrm{C/O}>1$ and
$\mathrm{\mathrm{^{12}C/^{13}C}<10}$ - properties characteristic
to J-stars. We identify systems which merge (COAL or CE) as a result
of unstable (dynamical or thermal) mass-transfer following a nova
phase during which the material ejected from the WD pollutes its companion
with enough carbon (including $\mathrm{^{13}C}$) to have $\mathrm{\mathrm{^{12}C/^{13}C}<10}$.
\\
The post-nova mergers evolve as AGB J-stars with a low-mass WD as
its core and the companion absorbed into its envelope. Such low mass
($<2\, M_{\odot}$) AGB stars have low $\mathrm{^{12}C/^{13}C}$ and
are dimmer than $M\mathrm{_{Bol}}=-5$, resembling observed J-stars
in the LMC that cannot otherwise be explained with single star (HBB)
models. However, our population synthesis study shows that this channel
together with HBB (single or binary) stars cannot account for the
observed fraction of J-stars among C-stars in the LMC and our Galaxy.
It indicates that other channels for the origin of J-stars need to
be taken into account in future studies aiming to explain the observed
frequency of J-stars. The GB-MS (Case B RLOF) mergers that possibly
behave differently from normal (single) AGB stars are of particular
interest and detailed evolutionary calculations are necessary to determine
their evolutionary time-scales. Other binary scenarios such as HeWD-giant
mergers (\citealt{2013MNRAS.430.2113Z}) should also be investigated
in a population synthesis framework to explain the observed fraction
of J-stars among C-stars. Such scenarios could also lead to possible
evolutionary connections with other classes of C-stars e.g. CEMP-no
stars which show chemical peculiarities common to J-stars at much
lower metallicities associated with the Galactic halo.
\begin{acknowledgements}
We acknowledge the anonymous referee for useful comments and interesting
suggestions for further studies. SS thanks K. Ohnaka for sharing his
observational expertise throughout the course of this work and for
his valuable comments to improve the manuscript. Special thanks also
go to J. Jose and R. J. Stancliffe for providing their models for
this work. SS received the financial support from the Bonn-Cologne
Graduate School for Physics and Astronomy and the Bonn International
Graduate School and is a member of the International Max Planck Research
School for Astronomy and Astrophysics at the Universities of Bonn
and Cologne. RGI and HHBL thank the Alexander von Humboldt foundation
for funding their positions in Bonn. 
\end{acknowledgements}
\bibliographystyle{aa}
\bibliography{/users/sutirtha/Documents/sutirtha/thesis/Jstars/Jstars,/users/sutirtha/Documents/J_stars/izzard/263/doc/references}

\appendix

\section{Constraining CE ejection efficiency \label{sec:Constraining-CE-ejection}}

Our choice of the CE ejection efficiency parameter $\alpha\mathrm{_{CE}}$
is based on the following population synthesis estimate for the Galactic
nova rate (per year) which is constrained by observations \citep{2002AIPC..637..462S}.
In the population synthesis algorithm for counting novae, the total
frequency ($\mathrm{\nu_{tot}}$) of nova-explosions (per year) for
$Z=0.02$ and a constant star formation rate, $S=7.086\,\mbox{\ensuremath{\mathrm{yr^{-1}}}}$
(following the prescription of \citealt{2002MNRAS_329_897H}), is
estimated from,

\begin{equation}
\nu_{\mathrm{tot}}=\sum_{i}S\times\delta p_{i}\times\left(\frac{\delta t_{\mathrm{nova}}}{\tau_{\mathrm{rec}}}\right)\mathrm{,}\label{eq:nova frequency}
\end{equation}
with the nova recurrence time $\mathrm{\tau{}_{rec}}$ for a binary
system calculated for every timestep $\mathrm{\delta t_{nova}}$ of
its evolution in a nova-phase from,

\begin{equation}
\tau\mathrm{_{rec}}=\frac{\Delta M_{\mathrm{WD}}^{\mathrm{crit}}}{\dot{M}}\,,\label{eq:recurrence times}
\end{equation}
using the expression of \citet{1995ApJ...447..656Y} for the critical
ignition mass $\Delta M_{\mathrm{WD}}^{\mathrm{crit}}$, 
\begin{equation}
\triangle M_{\mathrm{WD}}^{\mathrm{crit}}=2\times10^{-6}\left(\frac{M_{\mathrm{WD}}}{R_{\mathrm{WD}}^{4}}\right)^{0.8}\,.
\end{equation}
The WD radius $R\mathrm{_{WD}}$ is determined according to the formula
of \citet{1972ApJ...175..417N},

\begin{equation}
R_{\mathrm{WD}}=0.0112\sqrt{\left(\frac{M_{\mathrm{Ch}}}{M_{\mathrm{WD}}}\right)^{\frac{2}{3}}-\left(\frac{M_{\mathrm{Ch}}}{M_{\mathrm{WD}}}\right)^{-\frac{2}{3}}},
\end{equation}
where $M\mathrm{_{\mathrm{Ch}}=1.43\,\mathrm{M}_{\odot}}$ is the
Chandrasekhar mass.\\

\subsection{Galactic nova rate }

With the above prescription, the total rate of novae in the Galaxy
($Z=0.02$) is calculated using \textit{binary\_c/nucsyn} as,
\begin{equation}
R\mathrm{_{nova}=}f_{\mathrm{bin}}\times\nu_{\mathrm{tot}}\,,\label{eq:galactic novarate}
\end{equation}
assuming a constant binary fraction $f_{\mathrm{bin}}=0.5$ \citep{1991A&A...248..485D}.
As shown in Table~\ref{tab:Dependence-of-novae}, our estimate of
the Galactic nova rate is consistent with existing estimates of $\mathrm{30\pm10}\,\mbox{\ensuremath{\mathrm{yr^{-1}}}}$
\citep{2002AIPC..637..462S} within a factor of $\sim2-3$, being
closer to the observed rate for values of $\mathrm{\alpha_{CE}}<0.5$,
and decreasing with higher values of $\mathrm{\alpha_{CE}}$. 
\begin{table}[h]
\begin{centering}
\begin{tabular}{|c|c|c|c|}
\hline 
$\mathrm{\alpha{}_{CE}}$ & $R\mathrm{_{nova}}=f_{\mathrm{bin}}\times\nu_{\mathrm{tot}}$(yr$^{-1}$) & $\alpha$ & $\left\langle M_{\rm{WD}}\right\rangle$\tabularnewline
\hline 
\hline 
$0.1$ & $26.5$ & $0.021$ & $0.92$\tabularnewline
\hline 
$0.2$ & $49$ & $0.021$ & $1.07$\tabularnewline
\hline 
$0.3$ & $29.3$ & $0.02$ & $0.995$\tabularnewline
\hline 
$0.4$ & $18.6$ & $0.02$ & $0.92$\tabularnewline
\hline 
$0.5$ & $13.6$ & $0.019$ & $0.82$\tabularnewline
\hline 
$1.0$ & $10.9$ & $0.02$ & $0.77$\tabularnewline
\hline 
\end{tabular}
\par\end{centering}

\caption{\label{tab:Dependence-of-novae}Estimated values of the Galactic nova
rate, $R\mathrm{_{nova}}$, average WD-mass in nova systems, $<M\mathrm{_{WD}}>$,
and fraction of WD binaries leading to novae, $\alpha$ (according
to Eq.~\ref{eq:alpha}), as a function of the CE ejection efficiency
$\mathrm{\alpha_{CE}}$. The frequency of novae, $\nu_{\mathrm{tot}}$,
is calculated according to Eq.~ \ref{eq:nova frequency} and a constant
binary fraction $f_{\mathrm{bin}}=0.5$ is used to estimate $R\mathrm{_{nova}}$
according to Eq.~\ref{eq:galactic novarate}.}
\end{table}

\begin{figure}
\begin{raggedright}
\includegraphics[angle=270,scale=0.4]{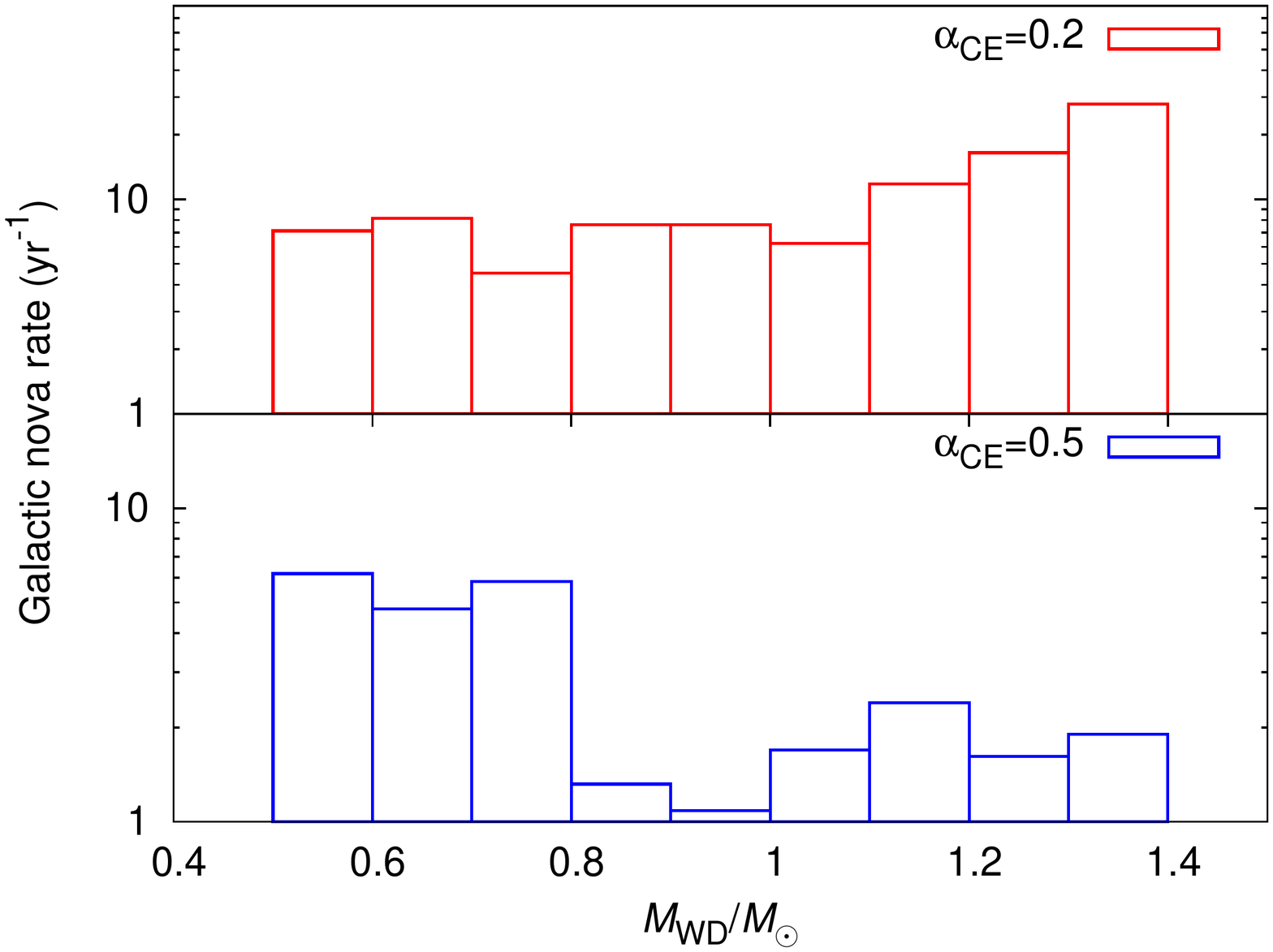}
\par\end{raggedright}

\caption{\label{fig:Histograms-of-predicted novarates}Histograms of predicted
Galactic nova rate as function of WD mass, $M\mathrm{_{WD}}$ per
$0.1\,\mbox{\ensuremath{M}}_{\odot}$ bin, for $\mathrm{\alpha_{CE}}=0.2\,\mbox{(upper panel) and }0.5\,$(lower
panel). The total rate of novae in the Galaxy $R\mathrm{_{nova}}$
is the sum of the contributions from all $M_{\mathrm{WD}}$ bins.}
\end{figure}

Fig.~\ref{fig:Histograms-of-predicted novarates} further illustrates
the sensitivity of the nova rate to the choice of $\mathrm{\alpha_{CE}}$.
The nova rate decreases sharply at higher WD masses and for more efficient
CE ejection. This is expected because the final separation after the
common envelope phase increases with $\mathrm{\alpha_{CE}}$, so that
for a higher $\mathrm{\alpha_{CE}}$, the post-CE binary is too wide
for the secondary to evolve to fill its Roche lobe. Because of shorter
recurrence times for massive WDs which lead to more frequent outbursts,
the total nova rate drops sharply with increasing $\mathrm{\alpha_{CE}\,}(0.2\mbox{ to }0.5)$.
\citet{2004ApJ...602..938N} also find an increase in the nova rate
by a factor of $\sim2-3$ for lower CE ejection efficiency. For $\mathrm{\alpha_{CE}}<0.2$,
the nova rate decreases as more systems merge following the CE phase
that leads to the WD-MS binary.

\subsection{Additional tests for nova binaries}

The fraction, $\alpha$, of binary systems hosting WDs that lead to
novae is also estimated as,
\begin{equation}
R_{\mathrm{nova}}=\alpha R_{\mathrm{WD}}^{\mathrm{bin}}\,,\label{eq:alpha}
\end{equation}
where $R_{\mathrm{WD}}^{\mathrm{bin}}$ is the birth rate of WDs in
binary systems. Also, the frequency averaged WD mass ($\left\langle M_{\rm{WD}}\right\rangle$
in $M_{\odot}$) is calculated for different choices of $\alpha_{\mathrm{CE}}\,(<1)$.
As shown in Table~\ref{tab:Dependence-of-novae}, our estimates for
this fraction $\alpha$ (defined by Eq.~\ref{eq:alpha}) is close
to the value of $0.02$ used by \citet{2005NuPhA.758..328R} to match
the observed nucleosynthesis yields for Galactic novae. The averaged
(by nova frequency) WD mass $\left\langle M_{\rm{WD}}\right\rangle$
is also in agreement with the values in the range of $0.8-1.1\, M_{\odot}$
as obtained by \citet{2004ApJ...602..938N}.

\section{HBB J-stars: single stars vs binary mergers \label{sec:HBB-J-stars:-Case-B}}

\begin{figure}[H]
\begin{centering}
\includegraphics[angle=270,scale=0.26]{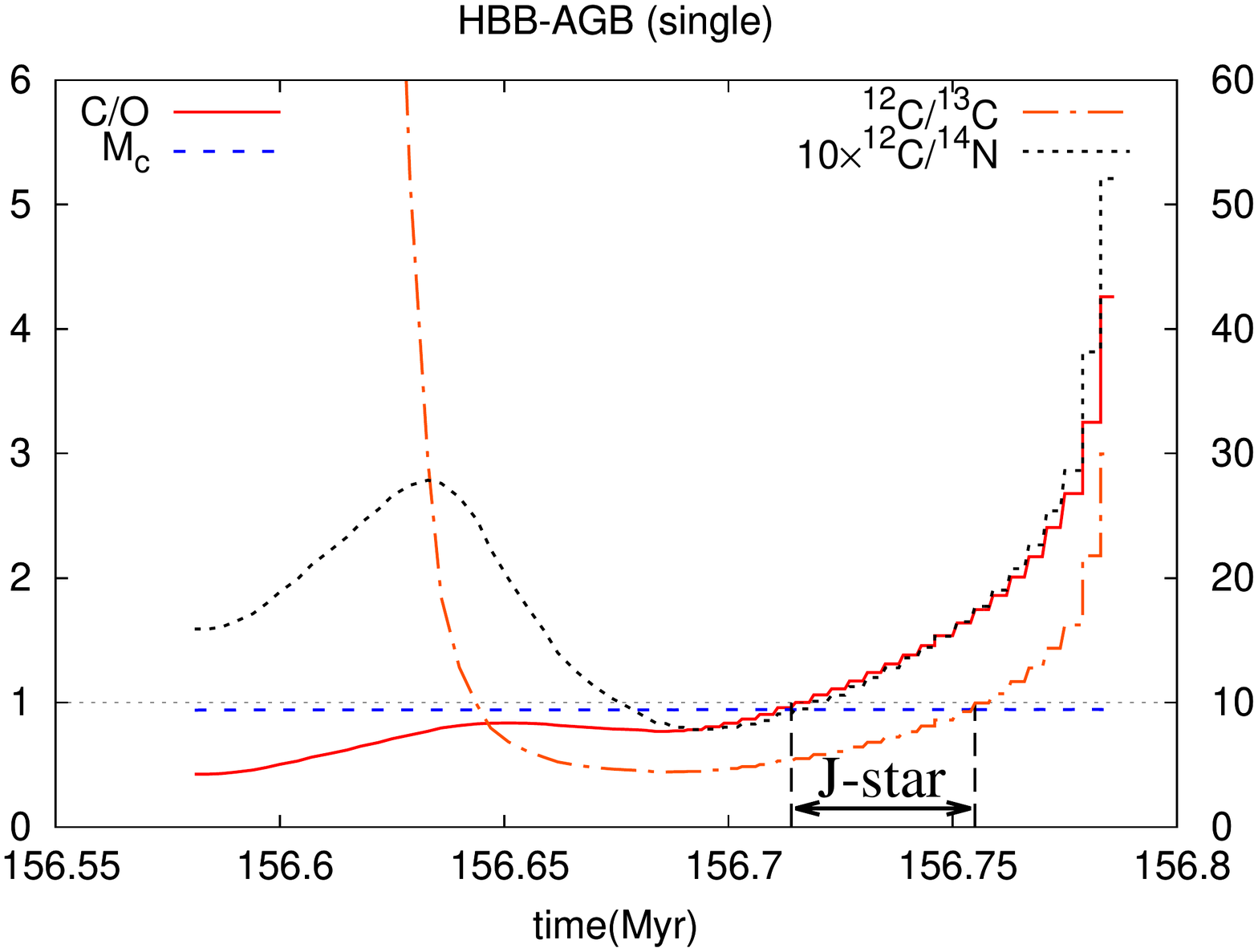}
\par\end{centering}

\begin{centering}
\includegraphics[angle=270,scale=0.26]{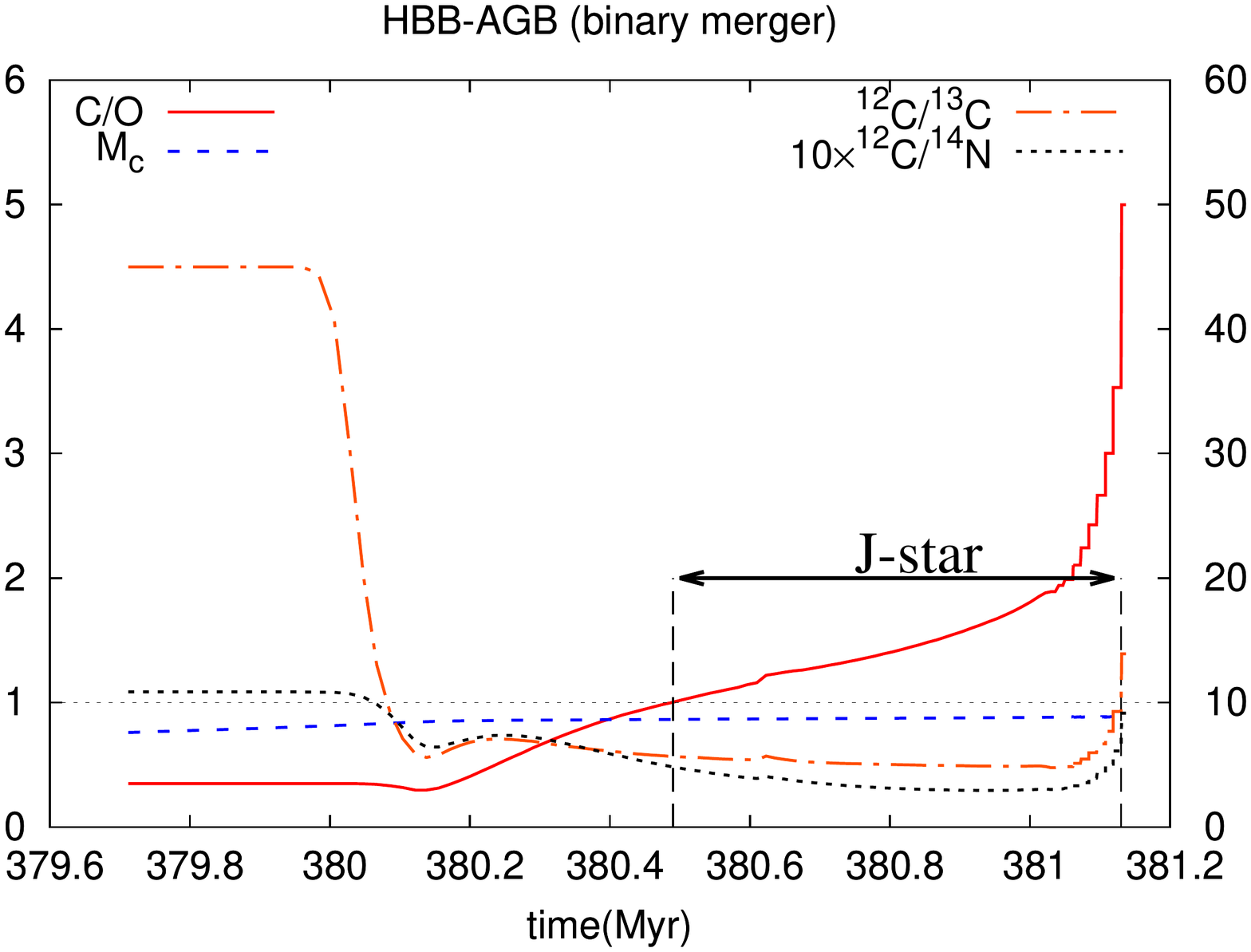}
\par\end{centering}

\caption{\label{fig:TPAGB-phase}Evolutionary properties on the TPAGB for a
single star (upper panel) and a GB-MS (Case-B RLOF) binary merger
(lower panel) of similar total mass. The core mass $M_{{\rm c}}$
and surface $\mathrm{C/O}$ are plotted on the left scale, along with
surface isotopic ratios, $\mathrm{\mathrm{^{12}C/^{13}C}\mbox{ and }{}^{12}C/^{14}N}$
(multiplied by a factor of $10$) on the right scale. The dotted horizontal
lines mark $\mathrm{C/O}=1$ and ${\rm ^{12}C/^{13}C=10}$ while the
dashed vertical lines mark the duration of the J-star phase.}
\end{figure}

If mass ratio $q$ exceeds a critical value $q_{{\rm crit}}$ (as
defined in \citealt{2002MNRAS_329_897H}), RLOF from a GB star to
a lower mass MS companion is dynamically unstable and mass transfer
leads to a CE phase in which the system can merge depending on the
initial separation. For such a Case-B RLOF merger, the MS star is
absorbed into the envelope and the core-mass of the merger is determined
by the core of the GB donor star at the onset of the CE phase. Consequently,
following the CE phase, the merged star has a lower core mass on the
GB compared to a single star of same total mass. Because the core
mass at the base of the AGB phase in our synthetic models depends
on the core mass at the base of the GB (\citealt{2002MNRAS_329_897H}),
such mergers ascend the AGB with lower core masses than single AGB
stars of similar total masses. Consequently they live longer on the
AGB and, for total masses higher than about $4\, M_{\odot}$ for which
HBB occurs along with TDU, we predict such mergers also behave as
J-stars for a longer phase on the AGB as compared to single HBB AGB
stars. Fig.~\ref{fig:TPAGB-phase} shows an example of the J-star
phase for such a binary merger that lasts for about $\mathrm{1Myr}$
compared to the much shorter span of about $\mathrm{0.1Myr}$ for
an equivalent single star. The binary system initially consists of
a $3.1\, M_{\odot}$ primary star which overflows its Roche lobe on
the GB with a core mass of $0.45\, M_{\odot}$ that subsequently forms
the core of the merged star following a CE phase with the $1.3\, M_{\odot}$
(MS) secondary absorbed into the giant envelope. Thus, the merger
core (and envelope) mass is significantly different from a single
star of similar total mass, and consequently the core mass can only
grow to $0.47\, M_{\odot}$ at the end of the GB - significantly lower
than the expected core mass ($\sim0.73\, M_{\odot}$) of a corresponding
single star.\textbf{ }This in turn leads to a lower core mass at the
start of the TPAGB for the binary merger and consequently it evolves
for a longer phase with TDU because of which its surface $\mathrm{C/O}$
exceeds $1$, and HBB that decreases the isotopic ratios of $\mathrm{\mathrm{^{12}C/^{13}C}\mbox{ and }{}^{12}C/^{14}N}$
on the stellar surface classifying it as a J-star.

\end{document}